\documentclass[12pt]{article}
\usepackage{amsmath,amssymb,amsthm,amsxtra,overpic,bbm,bm,epsfig,subfigure}
\usepackage{hyperref}
\usepackage{mathrsfs}
\usepackage{enumitem}
\usepackage{graphicx}
\usepackage{color}
\usepackage[table]{xcolor}
\usepackage{comment}
\usepackage{epstopdf}
\usepackage{float}
\usepackage{cite}
\usepackage{slashed,stmaryrd}
\usepackage{youngtab}
\usepackage{makecell}

\usepackage{diagbox}
\usepackage{blkarray}

\usepackage[title]{appendix}

\allowdisplaybreaks[1]

\def\thefootnote{\fnsymbol{footnote}}

\usepackage{multirow}

\newcommand{\Op}{\mathcal{O}}
\newcommand{\Opr}{\mathcal{R}}
\newcommand{\loopf}{\frac{1}{16\pi^2 \varepsilon}}
\newcommand{\tr}[1]{{\rm Tr}\left(#1\right)}

\newcommand{\rmi}{{\rm i}}

\newcommand{\befune}[1]{ \mu \frac{{\rm d}#1}{{\rm d}\mu}}
\newcommand{\befun}{ \dot{C}}

\usepackage{ulem}

\begin{document}
	
	{\normalsize \flushright TUM-HEP 1475/23\\}
	\vspace{0.2cm}
	
%

\begin{center}
{\Large\bf Revisiting Renormalization Group Equations of the SMEFT Dimension-Seven Operators}
\end{center}

\vspace{0.2cm}

\begin{center}
{\bf  Di Zhang}~\footnote{E-mail: di1.zhang@tum.de} 
\\
\vspace{0.2cm}
{\small
Physik-Department, Technische Universität München, James-Franck-Straße, 85748 Garching, Germany}
\end{center}

\vspace{1.5cm}

\begin{abstract}
	In this work, we revisit the renormalization group equations (RGEs) of dimension-seven (dim-7) operators in the Standard Model effective field theory (SMEFT) resulting from mixing among dim-7 operators themselves by means of the background field method. Adopting a recently proposed physical basis for dim-7 operators, we achieve the explicit RGEs of all non-redundant dim-7 operators in the SMEFT for the first time. Together with those originating from the dim-5 and dim-6 operators, these results constitute the complete RGEs of dim-7 operators, and hence can be exploited to study full RG-running effects on some lepton- or baryon-number-violating processes involving dim-7 operators in the SMEFT, such as neutrino masses, neutrinoless double beta decay, meson and nucleon decays. We perform an analysis of the structure and perturbative power counting of the obtained one-loop anomalous dimension matrix, which is consistent with  a non-renormalization theorem and the naive dimension analysis. Additionally, a partial check on some results is carried out by means of different tools and quantum field gauges.
\end{abstract}


\def\thefootnote{\arabic{footnote}}
\setcounter{footnote}{0}
\newpage

\section{Introduction}

After the Higgs boson was discovered by  ATLAS and CMS experiments at the Large Hadron Collider in 2012~\cite{ATLAS:2012yve,CMS:2012qbp}, the last piece of the Standard Model (SM) is completed. However, it is still believed that the SM is not complete for its inability to account for tiny neutrino masses, dark matter candidate, and the matter-antimatter asymmetry of the Universe~\cite{ParticleDataGroup:2022pth,Xing:2020ijf}. Regarded as an effective field theory, the SM is usually extended either by introducing new particles, interactions and symmetries to constitute some more fundamental ultraviolet (UV) theories (e.g., grand united theories~\cite{Pati:1973uk,Pati:1974yy,Fritzsch:1974nn,Georgi:1974sy}) or oppositely by including all possible non-renormalization operators to construct the so-called SM effective field theory (SMEFT)~\cite{Buchmuller:1985jz,Grzadkowski:2010es}. The former approach is certainly intriguing while the latter one is model-independent and hence more practical at the present stage where we know little about new physics and its underlying dynamics. Moreover, the SMEFT has the ability to capture indirect consequences of all sorts of new heavy degrees of freedom, which are encoded in Wilson coefficients of non-renormalizable operators. In the SMEFT, all operators consist of the SM fields and preserve the SM gauge symmetry, the number of which increases rapidly with their mass dimension but can be counted by means of the Hilbert series~\cite{Henning:2015alf}. Physical bases of the SMEFT including all non-redundant operators have been constructed up to dimension-twelve~\cite{Buchmuller:1985jz,Grzadkowski:2010es,Ma:2019gtx,Aoude:2019tzn,Lehman:2014jma,Liao:2016hru,Liao:2019tep,Zhang:2023kvw,Murphy:2020rsh,Li:2020gnx,Durieux:2019eor,AccettulliHuber:2021uoa,Liao:2020jmn,Li:2020xlh,Harlander:2023psl} but they are not unique. Within a specific physical basis, one may indirectly search for new physics by mapping Wilson coefficients of those higher-dimensional operators onto various low-energy observables~\cite{Henning:2014wua,Brivio:2017vri,Isidori:2023pyp}.

The SMEFT is valid below a cut-off energy scale $\Lambda$, which is roughly the energy scale of new physics and higher than the electroweak scale, and the Wilson coefficients of dimension-$n$ (dim-$n$) operators are suppressed by $1/\Lambda^{n-4}$. For a given UV model extending the SM only by heavy degrees of freedom, one can match it onto the SMEFT at the cut-off scale and take advantage of the SMEFT renormalization group equations (RGEs) to get the corresponding parameters at a low-energy scale, which can be used to confront low-energy consequences. During this procedure, large logarithms resulting from multiple energy scales in the UV model are resummed systematically by the SMEFT RGEs and hence the perturbative convergence is well improved. For the SMEFT itself, RGEs also play a crucial role in the global analysis of experimental data obtained at different energy scales. Therefore, as a powerful tool for resummation and an essential bridge to link physics at different energy scales, the SMEFT RGEs have been extensively studied for the dim-5~\cite{Babu:1993qv,Chankowski:1993tx,Antusch:2001ck}, dim-6~\cite{Jenkins:2013zja,Jenkins:2013wua,Alonso:2013hga,Alonso:2014zka,Davidson:2018zuo,Wang:2023bdw}, dim-7~\cite{Liao:2016hru,Liao:2019tep,Chala:2021juk,Zhang:2023kvw}, and dim-8~\cite{AccettulliHuber:2021uoa,Chala:2021pll,DasBakshi:2022mwk,Helset:2022pde,DasBakshi:2023htx,Chala:2023jyx} operators. Some attempts to understand the structure of the anomalous dimension matrix (i.e., zero entries and the perturbative power order) have been successfully made~\cite{Alonso:2014rga,Elias-Miro:2014eia,Cheung:2015aba,Bern:2019wie,Craig:2019wmo,Cao:2023adc,Jenkins:2013sda,Liao:2017amb}, which turn out to be very helpful.

In this work, we revisit the RGEs for dim-7 operators in the SMEFT, more specifically, those from the mixing among dim-7 operators themselves~\footnote{Recently, RGEs for dim-7 operators induced by the dim-5 and dim-6 ones have been achieved in Ref.~\cite{Zhang:2023kvw}.}. This kind of contributions to baryon-number-violating (BNV) operators is first discussed in Ref.~\cite{Liao:2016hru} but some non-trivial flavor relations between these operators are not taken into account. Then such contributions are reconsidered together with those to baryon-number-conserving (BNC) operators in Ref.~\cite{Liao:2019tep}, where non-trivial flavor relations are factored into a physical basis for dim-7 operators. However, the RGEs for Wilson coefficients of dim-7 operators are not explicitly presented in Ref.~\cite{Liao:2019tep}, and instead all results are listed in the form of counterterms with some redundant contributions from the so-called flavor-blind basis~\cite{Liao:2019tep}. To get the explicit RGE for a specific dim-7 operator, one further has to pick up all non-redundant contributions to the corresponding Wilson coefficient and then derive the final result with the help of these collected contributions. This is extremely inconvenient and even limits applications of those results to some extent. Therefore, the motivation for this work is twofold. With a different calculation strategy, our results may provide a necessary and independent cross-check against the previous ones. More importantly, taking advantage of a suitable physical basis for dim-7 operators in the SMEFT, we derive explicit RGEs for all non-redundant dim-7 operators, which are still lacking in literature and can be directly exploited in phenomenological study.

We adopt the background field method (BFM)~\cite{Abbott:1980hw,Abbott:1981ke,Abbott:1983zw} to calculate all counterterms in $d=4-2\varepsilon$ dimension regularization with the modified minimal subtraction scheme. The BFM can largely simplify the renormalization in gauge sector and keep all intermediate calculations and surely final results gauge invariant since the effective action of background fields preserves gauge symmetries. We slightly modify the source code of {\sf Matchmakereft}~\cite{Carmona:2021xtq} to enable it to calculate a set of one-particle-irreducible (1PI) diagrams with single dim-7 operator insertion. Then, counterterms in a Green's basis for dim-7 operators~\cite{Zhang:2023kvw} can be acquired and then converted to those in a physical basis by applying equations of motion (EoMs) of the SM fields. This strategy has the power to make the whole calculation procedure clearer and more traceable, and furthermore the obtained counterterms in the Green's basis can be partially used for deriving RGEs of higher-dimensional operators. To achieve explicit RGEs, we adopt the physical basis for dim-7 operators in Ref.~\cite{Zhang:2023kvw}. This basis is slightly different from that in Ref.~\cite{Liao:2019tep} but more suitable for our purpose since flavor indices of all operators are not restricted and hence run over all flavors. We perform an analysis of the obtained anomalous dimension matrix. It shows that our results are fully consistent with the non-renormalization theorem~\cite{Cheung:2015aba} and perturbative power analysis~\cite{Jenkins:2013sda,Liao:2017amb}. However, some discrepancies between our results for counterterms and those in literature appear. We carry out a partial check on those discrepancies to ensure the correctness of results.

The rest of this paper is organized as follows. In Sec.~\ref{sec:smeft}, the SMEFT Lagrangian and BFM are briefly introduced. In Sec.~\ref{sec:cal}, we describe the calculation procedure and  collect all explicit RGEs for dim-7 operators after carrying out all calculations. Analyses of the achieved one-loop anomalous dimension matrix and a partial check on results are performed in Sec.~\ref{sec:analysis}. In Sec.~\ref{sec:UV}, an explicit UV model is briefly discussed to illustrate the relevance of the achieved RGEs for dim-7 operators. The main conclusions are summarized in Sec.~\ref{sec:con}. Finally, all wave-function renormalization constants and also counterterms in the Green's basis are collected in Appendices~\ref{app:wf} and \ref{app:cou}, respectively.

\section{The SMEFT Lagrangian and Background Field Method}\label{sec:smeft}

Because we focus on RGEs resulting from the mixing among dim-7 operators, other lower- and higher-dimensional operators will not be involved in calculations and can be ignored. The SMEFT Lagrangian with only dim-7 operators is given by
\begin{eqnarray}\label{eq:LSMEFT}
	\mathcal{L}^{}_{\rm SMEFT} = \mathcal{L}^{}_{\rm SM} + \left( \sum_i C^i_7 \Op^{(7)}_i + {\rm h.c.} \right) \;,
\end{eqnarray}
where $\mathcal{L}^{}_{\rm SM}$ denotes the SM Lagrangian and $\Op^{(7)}_i$ and $C^{i}_7$ are dim-7 operators and the corresponding Wilson coefficients with $i$ running over all dim-7 operators. The SM Lagrangian is
\begin{eqnarray}\label{eq:LSM}
	\mathcal{L}^{}_{\rm SM} &=& - \frac{1}{4} G^{A}_{\mu\nu} G^{A\mu\nu} - \frac{1}{4} W^I_{\mu\nu} W^{I\mu\nu} - \frac{1}{4} B^{}_{\mu\nu} B^{\mu\nu} + \left( D^{}_\mu H \right)^\dagger \left( D^\mu H \right) - m^2 H^\dagger H - \lambda \left( H^\dagger H \right)^2 
	\nonumber
	\\
	&& + \sum^{}_f \overline{f} \rmi \slashed{D} f - \left[ \overline{Q^{}_{\alpha \rm L}} \left(Y^{}_{\rm u} \right)^{}_{\alpha \beta} \widetilde{H} U^{}_{\beta \rm R} +  \overline{Q^{}_{\alpha \rm L}}  (Y^{}_{\rm d})^{}_{\alpha \beta} H D^{}_{\beta \rm R}+ \overline{\ell^{}_{\alpha \rm L}}(Y^{}_l)^{}_{\alpha \beta} H E^{}_{\beta \rm R}+ {\rm h.c.} \right]
	\nonumber
	\\
	&& + \mathcal{L}^{}_{\rm GF} + \mathcal{L}^{}_{\rm Ghost}   \;,
\end{eqnarray}
in which $f= Q^{}_{\rm L}, U^{}_{\rm R}, D^{}_{\rm R}, \ell^{}_{\rm L}, E^{}_{\rm R}$, the covariant derivative $D^{}_\mu \equiv \partial^{}_\mu - \rmi g^{}_1 Y B^{}_\mu - \rmi g^{}_2 T^I W^I_\mu - \rmi g^{}_3 T^A G^A_\mu$ is defined as usual, and $ \mathcal{L}^{}_{\rm GF} $ and $ \mathcal{L}^{}_{\rm Ghost}$  are respectively the gauge-fixing and Faddeev-Popov ghost terms depending on a chosen gauge. A set of ``independent" dim-7 operators, i.e., an operator basis has been first constructed in Ref.~\cite{Lehman:2014jma} but later two redundant operators and some redundancies due to non-trivial flavor relations have been removed to form a genuine physical basis~\cite{Liao:2016hru,Liao:2019tep}. Nevertheless, the physical basis given in Ref.~\cite{Liao:2019tep} is not so convenient for applications because flavor indices of some operators in that basis are restricted and symmetries among them are spoiled. To avoid such inconvenience, we adopt a slightly different basis proposed in Ref.~\cite{Zhang:2023kvw}, which keeps symmetries among flavor indices of all operators and is reproduced in Table~\ref{tab:phyb}. Note that compared with the basis in Ref.~\cite{Liao:2019tep}, operators from flavor combinations of $\Op^{}_{\overline{e}\ell\ell\ell H}$ and $\Op^{}_{\overline{\ell}dddH}$ are totally different, and $\Op^{}_{\ell HB}$ and $\Op^{}_{\ell HW}$ are without factors $g^{}_1$ and $g^{}_2/2$, respectively. These factors indeed matter when one derives RGEs of the corresponding Wilson coefficients.

\begin{table}
	\centering
	\renewcommand\arraystretch{1.8}
	\resizebox{\textwidth}{!}{
		\begin{tabular}{l|ll}
			\hline\hline
			\multicolumn{3}{c}{$\psi^2H^4$} \\
			\hline
			$\Op^{(S)\alpha\beta}_{\ell H} $ &  $ \displaystyle\frac{1}{2} \left( \Op^{\alpha\beta}_{\ell H} + \Op^{\beta\alpha}_{\ell H} \right) $ & with $\Op^{\alpha\beta}_{\ell H} = \epsilon^{ab} \epsilon^{de}  \left( \ell^a_{\alpha\rm L} C \ell^d_{\beta\rm L} \right) H^b H^e \left( H^\dagger H \right)$  \\
			\hline
			\multicolumn{3}{c}{$\psi^2H^3D$} \\
			\hline
			$\Op^{\alpha\beta}_{\ell eHD} $ & $\epsilon^{ab} \epsilon^{de} \left( \ell^a_{\alpha\rm L} C \gamma^{}_\mu E^{}_{\beta\rm R} \right) H^b H^d {\rm i} D^\mu H^e$  &  \\
			\hline
			\multicolumn{3}{c}{$\psi^2H^2 D^2$} \\
			\hline
			$ \Op^{(S)\alpha\beta}_{\ell HD1} $ & $ \displaystyle\frac{1}{2} \left( \Op^{\alpha\beta}_{\ell HD1} + \Op^{\beta\alpha}_{\ell HD1} \right) $  & with $\Op^{\alpha\beta}_{\ell HD1} = \epsilon^{ab} \epsilon^{de} \left( \ell^a_{\alpha\rm L} C D^\mu \ell^b_{\beta\rm L} \right) H^d D^{}_\mu H^e$ \\
			$ \Op^{(S)\alpha\beta}_{\ell HD2}  $ & $ \displaystyle\frac{1}{2} \left( \Op^{\alpha\beta}_{\ell HD2} + \Op^{\beta\alpha}_{\ell HD2} \right)$ & with $\Op^{\alpha\beta}_{\ell HD2} = \epsilon^{ad} \epsilon^{be} \left( \ell^a_{\alpha\rm L} C D^\mu \ell^b_{\beta\rm L} \right) H^d D^{}_\mu H^e$ \\
			\hline
			\multicolumn{3}{c}{$\psi^2H^2 X$} \\
			\hline
			$\Op^{(A)\alpha\beta}_{\ell HB} $ & $ \displaystyle\frac{1}{2} \left( \Op^{\alpha\beta}_{\ell HB} - \Op^{\beta\alpha}_{\ell HB} \right) $ & with $ \Op^{\alpha\beta}_{\ell HB} = \epsilon^{ab} \epsilon^{de} \left( \ell^a_{\alpha\rm L} C \sigma^{}_{\mu\nu} \ell^d_{\beta\rm L} \right) H^b H^e B^{\mu\nu}$  \\
			$\Op^{\alpha\beta}_{\ell HW} $ & $\epsilon^{ab} \left( \epsilon \sigma^I \right)^{de} \left( \ell^a_{\alpha\rm L} C \sigma^{}_{\mu\nu} \ell^d_{\beta\rm L} \right) H^b H^e W^{I \mu\nu}$ & \\
			\hline
			\multicolumn{3}{c}{$\psi^4H$} \\
			\hline
			$ \Op^{(S)\alpha\beta\gamma\lambda}_{\overline{e}\ell\ell\ell H} $ & $ \displaystyle\frac{1}{6} \left( \Op^{\alpha\beta\gamma\lambda}_{\overline{e}\ell\ell\ell H} + \Op^{\alpha\lambda\beta\gamma}_{\overline{e}\ell\ell\ell H} +  \Op^{\alpha\gamma\lambda\beta}_{\overline{e}\ell\ell\ell H} + \Op^{\alpha\beta\lambda\gamma}_{\overline{e}\ell\ell\ell H} + \Op^{\alpha\gamma\beta\lambda}_{\overline{e}\ell\ell\ell H} + \Op^{\alpha\lambda\gamma\beta}_{\overline{e}\ell\ell\ell H} \right) $ & with $\Op^{\alpha\beta\gamma\lambda}_{\overline{e}\ell\ell\ell H} = \epsilon^{ab} \epsilon^{de} \left( \overline{E^{}_{\alpha\rm R}} \ell^a_{\beta\rm L}\right) \left( \ell^b_{\gamma\rm L} C \ell^d_{\lambda\rm L} \right) H^e$  \\
			$\Op^{(A)\alpha\beta\gamma\lambda}_{\overline{e}\ell\ell\ell H} $ & $\displaystyle\frac{1}{6} \left( \Op^{\alpha\beta\gamma\lambda}_{\overline{e}\ell\ell\ell H} + \Op^{\alpha\lambda\beta\gamma}_{\overline{e}\ell\ell\ell H} +  \Op^{\alpha\gamma\lambda\beta}_{\overline{e}\ell\ell\ell H} - \Op^{\alpha\beta\lambda\gamma}_{\overline{e}\ell\ell\ell H} - \Op^{\alpha\gamma\beta\lambda}_{\overline{e}\ell\ell\ell H} - \Op^{\alpha\lambda\gamma\beta}_{\overline{e}\ell\ell\ell H} \right)$ & \\
			$\Op^{(M)\alpha\beta\gamma\lambda}_{\overline{e}\ell\ell\ell H} $ & $\displaystyle\frac{1}{3} \left( \Op^{\alpha\beta\gamma\lambda}_{\overline{e}\ell\ell\ell H} + \Op^{\alpha\gamma\beta\lambda}_{\overline{e}\ell\ell\ell H} -  \Op^{\alpha\lambda\gamma\beta}_{\overline{e}\ell\ell\ell H} - \Op^{\alpha\gamma\lambda\beta}_{\overline{e}\ell\ell\ell H} \right)$ &  \\
			$\Op^{\alpha\beta\gamma\lambda}_{\overline{d}\ell q \ell H1} $ & $\epsilon^{ab} \epsilon^{de} \left( \overline{D^{}_{\alpha\rm R}} \ell^a_{\beta\rm L} \right) \left( Q^b_{\gamma\rm L} C \ell^d_{\lambda\rm L} \right) H^e $ & \\ 
			$\Op^{\alpha\beta\gamma\lambda}_{\overline{d}\ell q \ell H2} $ & $\epsilon^{ad} \epsilon^{be} \left( \overline{D^{}_{\alpha\rm R}} \ell^a_{\beta\rm L} \right) \left( Q^b_{\gamma\rm L} C \ell^d_{\lambda\rm L} \right) H^e$ & \\ 
			$\Op^{\alpha\beta\gamma\lambda}_{\overline{d}\ell ueH} $ & $\epsilon^{ab} \left( \overline{D^{}_{\alpha\rm R}} \ell^a_{\beta\rm L} \right) \left( U^{}_{\gamma\rm R} C E^{}_{\lambda\rm R} \right) H^b$  & \\
			$\Op^{\alpha\beta\gamma\lambda}_{\overline{q} u \ell \ell H} $ & $\epsilon^{ab} \left( \overline{Q^{}_{\alpha\rm L}} U^{}_{\beta\rm R} \right) \left( \ell^{}_{\gamma\rm L} C \ell^a_{\lambda\rm L} \right) H^b$  & \\
			$\Op^{\alpha\beta\gamma\lambda}_{\overline{\ell}dud\widetilde{H}} $ & $\left( \overline{\ell^{}_{\alpha\rm L}} D^{}_{\beta\rm R} \right) \left( U^{}_{\gamma\rm R} C D^{}_{\lambda\rm R} \right) \widetilde{H} $ & \\
			$\Op^{(M)\alpha\beta\gamma\lambda}_{\overline{\ell} dddH} $  & $ \displaystyle\frac{1}{3} \left( \Op^{\alpha\beta\gamma\lambda}_{\overline{\ell} dddH} + \Op^{\alpha\gamma\beta\lambda}_{\overline{\ell} dddH} -  \Op^{\alpha\beta\lambda\gamma}_{\overline{\ell} dddH} - \Op^{\alpha\lambda\beta\gamma}_{\overline{\ell} dddH} \right)$ & with $\Op^{\alpha\beta\gamma\lambda}_{\overline{\ell} dddH} = \left( \overline{\ell^{}_{\alpha\rm L}} D^{}_{\beta\rm R} \right) \left( D^{}_{\gamma\rm R} C D^{}_{\lambda\rm R} \right) H$ \\
			$\Op^{(A)\alpha\beta\gamma\lambda}_{\overline{e}qdd\widetilde{H}}$ & $ \displaystyle\frac{1}{2} \left( \Op^{\alpha\beta\gamma\lambda}_{\overline{e}qdd\widetilde{H}} - \Op^{\alpha\beta\lambda\gamma}_{\overline{e}qdd\widetilde{H}}  \right) $  & with $\Op^{\alpha\beta\gamma\lambda}_{\overline{e}qdd\widetilde{H}}  = \epsilon^{ab} \left( \overline{ E^{}_{\alpha\rm R}} Q^a_{\beta\rm L} \right) \left( D^{}_{\gamma\rm R} C D^{}_{\lambda\rm R} \right) \widetilde{H}^b$ \\
			$ \Op^{\alpha\beta\gamma\lambda}_{\overline{\ell}dqq\widetilde{H}}$ & $\epsilon^{ab} \left( \overline{\ell^{}_{\alpha\rm L}} D^{}_{\beta\rm R} \right) \left( Q^{}_{\gamma\rm L} C Q^a_{\lambda\rm L} \right) \widetilde{H}^b$ & \\
			\hline
			\multicolumn{3}{c}{$\psi^4D$} \\
			\hline
			$\Op^{(S)\alpha\beta\gamma\lambda}_{\overline{e} dddD} $   &  $ \displaystyle\frac{1}{6} \left( \Op^{\alpha\beta\gamma\lambda}_{\overline{e}dddD} + \Op^{\alpha\lambda\beta\gamma}_{\overline{e}dddD} +  \Op^{\alpha\gamma\lambda\beta}_{\overline{e}dddD} + \Op^{\alpha\beta\lambda\gamma}_{\overline{e}dddD} + \Op^{\alpha\gamma\beta\lambda}_{\overline{e}dddD} + \Op^{\alpha\lambda\gamma\beta}_{\overline{e}dddD} \right)$  & with $\Op^{\alpha\beta\gamma\lambda}_{\overline{e}dddD} = \left(\overline{E^{}_{\alpha\rm R}} \gamma^{}_\mu D^{}_{\beta\rm R} \right) \left( D^{}_{\gamma\rm R} C {\rm i} D^\mu D^{}_{\lambda\rm R} \right) $ \\
			$\Op^{(S)\alpha\beta\gamma\lambda}_{\overline{d}u\ell\ell D}$ &  $ \displaystyle\frac{1}{2} \left( \Op^{\alpha\beta\gamma\lambda}_{\overline{d} u\ell\ell D} + \Op^{\alpha\beta\lambda\gamma}_{\overline{d} u\ell\ell D} \right) $ & with $\Op^{\alpha\beta\gamma\lambda}_{\overline{d} u\ell\ell D} = \epsilon^{ab} \left( \overline{D^{}_{\alpha\rm R}} \gamma^{}_\mu U^{}_{\beta\rm R} \right) \left( \ell^a_{\gamma\rm L} C {\rm i} D^\mu \ell^b_{\lambda\rm L} \right)$  \\ $\Op^{(S)\alpha\beta\gamma\lambda}_{\overline{\ell}qddD} $ & $ \displaystyle\frac{1}{2} \left( \Op^{\alpha\beta\gamma\lambda}_{\overline{\ell} qdd D} + \Op^{\alpha\beta\lambda\gamma}_{\overline{\ell} qdd D} \right)$ & with $\Op^{\alpha\beta\gamma\lambda}_{\overline{\ell} qdd D}  = \left( \overline{\ell^{}_{\alpha\rm L}} \gamma^{}_\mu Q^{}_{\beta\rm L} \right) \left( D^{}_{\gamma\rm R} C {\rm i} D^\mu D^{}_{\lambda\rm  R} \right)$  \\
			\hline\hline
		\end{tabular}
	}
	\caption{A physical basis for dim-7 operators adopted in this work and taken from Ref.~\cite{Zhang:2023kvw}.}\label{tab:phyb}
\end{table}

To fix the gauge-fixing and ghost field terms in Eq.~\eqref{eq:LSM}, the quantization of the SM has to be performed. If one quantizes a gauge theory in the conventional formulation, the gauge fixing term will spoil the gauge symmetry. As a result, gauge invariance is not preserved in intermediate steps of calculations and can only be restored at the very end. Fortunately, it is proved that the BFM can be used to avoid such an explicit breaking of gauge symmetry and hence largely simplify calculations~\cite{Abbott:1980hw,Abbott:1981ke}. Working with the BFM, one splits each field into background and quantum parts, and the former only appears as external lines or internal tree-level propagators while the latter as internal loop propagators. Then, a proper gauge condition can be chosen to guarantee that the background field effective action has explicit gauge invariance with respect to gauge transformations of background fields. This indicates that all 1PI Green's functions generated by the effective action and also divergent counterterms have a gauge invariant form. Moreover, the naive Ward identities of gauge invariance hold in this framework, which can simplify the renormalization of gauge sector as we will see later. The gauge condition preserving gauge symmetry for background fields is found to be
\begin{eqnarray}\label{eq:gc}
	\mathcal{G}^i_V = D^{}_\mu \hat{V}^{i\mu}= \partial_\mu\hat {V}^{i\mu} - \rmi g_V T^j_{ik} V^j_\mu \hat{V}^{k\mu} \;,
\end{eqnarray}
where $V = B,W,G$ (or $\hat{V} = \hat{B},\hat{W},\hat{G}$) are background (or quantum) gauge fields, $g_V$ and $T^j$ with $j$ being adjoint index are respectively the gauge coupling and the adjoint representation of the corresponding gauge group. With the gauge condition in Eq.~\eqref{eq:gc}, the gauge fixing and ghost field terms in Eq.~\eqref{eq:LSM} turn out to be
\begin{eqnarray}\label{eq:LGF}
	\mathcal{L}^{}_{\rm GF} &=&  - \frac{1}{2\xi_B} \left( \partial^\mu \hat{B}^{}_\mu \right)^2 - \frac{1}{2\xi_W}  \left( D^\mu \hat{W}^I_{\mu} \right)^2 - \frac{1}{2\xi_G}  \left( D^\mu \hat{G}^A_{\mu} \right)^2 \;,
	\nonumber
	\\
	\mathcal{L}^{}_{\rm Ghost} &=& - \overline{\theta}^{}_{B} \partial^{}_\mu \partial^\mu \theta^{}_{B} - \overline{\theta}^{I}_{W} D^{}_\mu \hat{D}^\mu \theta^{I}_{W} - \overline{\theta}^{A}_{G} D^{}_\mu \hat{D}^\mu \theta^{A}_{G} \;,
\end{eqnarray}
where $\xi^{}_{V}$ and $\theta^{}_V$ (for $V=B,W,G$) denote gauge parameters and ghost fields for the corresponding gauge fields $\hat{V}$ in the general $R^{}_\xi$ gauge, and $\hat{D}^{}_\mu$ stands for the covariant derivative with both background and quantum gauge fields, i.e., substituting $V^{}_\mu+\hat{V}^{}_\mu$ for $V^{}_\mu$ in the background covariant derivative $D^{}_\mu$. Note that the first term in $\mathcal{L}^{}_{\rm Ghost}$ only  involves ghost fields and can be omitted. In principle, apart from those in gauge fixing and ghost field terms given in Eq.~\eqref{eq:LGF}, all fields in Eq.~\eqref{eq:LSM} and also those in dim-7 operators should be replaced with the corresponding background and quantum fields. Fortunately, with the chosen gauge condition in Eq.~\eqref{eq:gc}, fermion and Higgs fields are not drawn into the gauge fixing term in Eq.~\eqref{eq:LGF}. This means that the background and quantum fields for fermions or Higgs have exactly the same interactions and there is no need to distinguish them. Thus, one can keep fermion and Higgs fields in the Lagrangian unchanged and only split gauge fields into background and quantum parts. Furthermore, if one is only concerned with 1PI Green's function, it is not necessary to choose a gauge fixing term for background gauge fields since they only appear as external lines in 1PI diagrams. 

As pointed out in Ref.~\cite{Abbott:1980hw}, we do not need to renormalize quantum and ghost fields when performing renormalization with the BFM, because these fields always appear in loops and renormalization constants associated with propagators and the adjacent vertices are fully canceled out. Nevertheless, gauge fixing parameters $\xi^{}_V$ for quantum gauge fields $\hat{V}=\hat{B},\hat{W},\hat{G}$ still need to be renormalized. Fortunately, they are not relevant for the one-loop renormalization. On the other hand, the renormalization of background gauge parts with the BFM is much simpler than the conventional one due to gauge invariance of the background field effective action, but those of fermion and Higgs sectors keep the same as the conventional one. More specifically, the renormalization constants $Z^{}_{g_V}$ for gauge couplings $g^{}_V$ are fully related to wave function renormalization constants $Z^{}_V$ of background gauge fields $V=B,W,G$, namely $Z^{}_{g_V} = Z^{-1/2}_V$. This means that the self-energy of background gauge fields is enough for the renormalization of background gauge sector. Thanks to its gauge invariance and simplification of renormalization, the BFM has been extensively applied to matchings of UV models onto the SMEFT and calculations of RGEs in the SMEFT.

\section{Calculations and RGEs for Dim-7 Operators}\label{sec:cal}

We make use of the Green's and physical bases for dim-7 operators put forward in Ref.~\cite{Zhang:2023kvw}. The former one is given in Appendix~\ref{app:cou} and the latter is shown in Table.~\ref{tab:phyb}. The calculation procedure is briefly summarized as below:
\begin{itemize}
	\item Generating all 1PI diagrams with single insertion of the physical dim-7 operators in Table~\ref{tab:phyb}, whose external lines are determined by the explicit field ingredients of a specific operator class in the Green's basis, such as $\left\langle \ell\ell HH \right\rangle$ for $\psi^2 H^2 D^2$ class.
	\item Calculating the generated 1PI diagrams to extract UV divergences and also contributions from the corresponding counterterms in the Green's basis, e.g., $\delta G^{}_{\ell HDi} \Op^{}_{\ell HDi}$ (for $i=1,2,...,6$) for $\left\langle \ell\ell HH \right\rangle$~\footnote{The dim-5 operator is also needed to cancel out all UV divergences.}.
	\item Solving a group of linear equations to obtain explicit expressions for counterterms. These linear equations result from the requirement of UV divergence cancellation.
	\item Repeating the above procedures to cover field ingredients of all operator classes in the Green's basis. Then, one achieves counterterms for all operators in the Green's basis.
	\item Converting all results in the Green's basis to those in the physical basis by means of the reduction relations as a result of EoMs.
	\item Calculating RGEs for Wilson coefficients of dim-7 operators from those counterterms in the physical basis.
\end{itemize}

The whole calculation is quite lengthy and involved owing to a plethora of relevant 1PI diagrams. Fortunately, all relevant diagrams can be generated and calculated (i.e., the first two steps described above) by means of the {\sf Matchmakereft} package~\cite{Carmona:2021xtq} with the Feynman-'t Hooft gauge $\xi^{}_{V} = 1$ (for $V=B,W,G$). But one has to modify some original codes of {\sf Matchmakereft} to make the package competent in doing those calculations and giving results in a right form. For instance, the package can only deal with pour four-fermion vertices, but not those involving some additional fields besides four fermion fields so far. This means the original package can not handle any diagram containing one dim-7 operator in $\psi^4 H$ class unless the WriteVertices function in FR2MM.m file is modified properly~\footnote{The modified files can be found in the ancillary files on arXiv.org.}. On the other hand, not all Fierz identities are applied in the package, thus fermion chains in the output are not fully independent. Before figuring out explicit expressions for counterterms in the third step, one must make use of some necessary Fierz identities to refine the output. One can build those Fierz identities into the package, but we decide to directly apply them to the output. In addition to those explicitly listed in Appendix A of Ref.~\cite{Liao:2016hru}, some generalized Fierz identities involving simultaneously vector and tensor fermionic currents are indispensable as well, namely~\cite{Nieves:2003in,Nishi:2004st,Liao:2012uj}, 
\begin{eqnarray}\label{eq:gfi}
	\left( \overline{\psi^{}_{1\rm L}} \gamma^{}_\mu \psi^{}_{2\rm L} \right) \left( \overline{\psi^{}_{3\rm R}} \sigma^{\mu\nu} \psi^{}_{4 \rm L} \right) &=& - \frac{3}{2}\rmi \left( \overline{\psi^{}_{1\rm L}} \gamma^\nu \psi^{}_{4\rm L} \right) \left( \overline{\psi^{}_{3\rm R}} \psi^{}_{2\rm L} \right) + \frac{1}{2}  \left( \overline{\psi^{}_{1\rm L}} \gamma^{}_\mu \psi^{}_{4\rm L} \right) \left( \overline{\psi^{}_{3\rm R}} \sigma^{\mu\nu} \psi^{}_{2\rm L} \right) \;,
	\nonumber
	\\
	\left( \overline{\psi^{}_{1\rm L}} \gamma^{}_\mu \psi^{}_{2\rm L} \right) \left( \overline{\psi^{}_{3\rm L}} \sigma^{\mu\nu} \psi^{}_{4 \rm R} \right) &=& \frac{3}{2}\rmi \left( \overline{\psi^{}_{1\rm L}}  \psi^{}_{4\rm R} \right) \left( \overline{\psi^{}_{3\rm L}} \gamma^\nu \psi^{}_{2\rm L} \right) + \frac{1}{2}  \left( \overline{\psi^{}_{1\rm L}} \sigma^{\mu\nu} \psi^{}_{4\rm R} \right) \left( \overline{\psi^{}_{3\rm L}} \gamma^{}_\mu \psi^{}_{2\rm L} \right) \;,
\end{eqnarray}
that are also valid for exchanging left- and right-handed chiralities of all fermion fields, i.e., $\rm L \leftrightarrow \rm R$. The above Fierz identities play an important role in calculating counterterms for dim-7 operators in $\psi^4 D$ class in the Green's basis. One can see that operators with tensor fermionic current in $\psi^4 D$ class can be reduced to other operators without tensor fermionic current in the same class. This is the reason why there could be no operator with tensor fermionic current in $\psi^4 D$ class in the Green's basis as shown in Table~\ref{tab:green-basis}.  Here, we simply give an example, that is
\begin{eqnarray}
	\epsilon^{ab} \left( \overline{D^{}_{\alpha\rm R}} \gamma^{}_\mu U^{}_{\beta\rm R} \right) \left( \ell^a_{\gamma\rm L} C \sigma^{\mu\nu} D^{}_\nu \ell^b_{\lambda\rm L} \right) &=&-\frac{3}{2} \Opr^{\alpha\lambda\gamma\beta}_{\overline{d}D\ell\ell u}  + \frac{1}{2} \epsilon^{ab} \left( \overline{D^{}_{\alpha\rm R}} \sigma^{\mu\nu} D^{}_\nu \ell^b_{\lambda\rm L} \right) \left( \ell^a_{\gamma\rm L} C \gamma^{}_\mu U^{}_{\beta\rm R} \right) 
	\nonumber
	\\
	&=&  -\frac{3}{2} \Opr^{\alpha\lambda\gamma\beta}_{\overline{d}D\ell\ell u} + \frac{3}{4} \Op^{\alpha\beta\gamma\lambda}_{\overline{d}u\ell\ell D} + \frac{1}{4} \epsilon^{ab} \left( \overline{D^{}_{\alpha\rm R}} \gamma^{}_\mu U^{}_{\beta\rm R} \right) \left( \ell^a_{\gamma\rm L} C \sigma^{\mu\nu} D^{}_\nu \ell^b_{\lambda\rm L} \right) 
	\nonumber
	\\
	&=& \Op^{\alpha\beta\gamma\lambda}_{\overline{d}u\ell\ell D} - 2 \Opr^{\alpha\lambda\gamma\beta}_{\overline{d}D\ell\ell u} \;,
\end{eqnarray}
where the generalized Fierz identities in Eq.~\eqref{eq:gfi} have been used twice. With the original output of {\sf Matchmakereft} and applying necessary Fierz identities, one may acquire a set of linear equations for each group of operators with the same explicit field ingredients via requiring all UV divergences to be canceled out by the corresponding counterterms. Then, explicit results for all counterterms in the Green's basis can be achieved by solving those linear equations. For our purpose, wave-function renormalization constants of the SM fields are also essential, but the SM version for them is already enough, which are simultaneously calculated by {\sf Matchmakereft} and included in the output. We list all wave-function renormalization constants and all counterterms in the Green's basis in Appendices~\ref{app:wf} and \ref{app:cou} for the convenience of deriving higher-dimensional operators' RGEs and for crosschecking results as well. To obtain results in the physical basis, the EoMs of the SM fields should be implemented, or equivalently, one can directly take advantage of the reduction relations for dim-7 operators in the adopted Green's and physical bases, which exactly result from the EoMs and have been worked out in Ref.~\cite{Zhang:2023kvw}. Finally, one can derive RGEs for Wilson coefficients of all dim-7 operators from the corresponding counterterms and wave-function renormalization constants with the help of the following formula:
\begin{eqnarray}\label{eq:grge}
	\befune{C^{}_r} = \varepsilon \left( \sum^{}_i n^\prime_i h^{}_i \frac{\partial Z^{}_r}{\partial h^{}_i} \right) C^{}_r
\end{eqnarray}
where $C^{}_r$ is the renormalized Wilson coefficient, $h^{}_i$ runs over all couplings including Wilson coefficients, and $n^\prime_i$ is the tree-level anomalous dimension for the corresponding coupling. The renormalization constant $Z^{}_r$ in Eq.~\eqref{eq:grge} is given by
\begin{eqnarray}\label{eq:ct-wf}
	Z^{}_r &=& 1 - \sum^{}_{\varphi} \frac{1}{2} n^{}_\varphi \delta Z^{}_\varphi + \frac{\delta C^{}_r}{C^{}_r} \;,
\end{eqnarray}
in which $\delta Z^{}_\varphi$ and $n^{}_\varphi$ are respectively the wave-function renormalization constant and the number of the field $\varphi$ appearing in the corresponding operator, and $\delta C^{}_r$ is the counterterm of $C^{}_r$. For simplicity, flavor structures are not taken into account in Eqs.~\eqref{eq:grge} and \eqref{eq:ct-wf}, but it is easy to generalize the above formulae for Wilson coefficients with flavor indices.

In the rest of this section, we collect all results for RGEs of Wilson coefficients organized by operator classes, where $\dot{C} \equiv16\pi^2 \mu {\rm d} C/{\rm d} \mu$ and $T \equiv \tr{ Y^{}_l Y^\dagger_l + 3 Y^{}_{\rm u} Y^\dagger_{\rm u} + 3 Y^{}_{\rm d} Y^\dagger_{\rm d} } $ have been introduced for brevity, and $\alpha \leftrightarrow \beta$ stands for exchanging flavors $\alpha$ and $\beta$ for all terms.
\\
\\
\noindent$\bullet~\bm{\psi^2H^4}$
\begin{eqnarray}\label{eq:psi2H4}
	\befun^{(S)\alpha\beta}_{\ell H} &=& - \frac{1}{4} \left( 3g^2_1 + 15g^2_2 - 80 \lambda - 8 T \right) C^{(S)\alpha\beta}_{\ell H} - \frac{3}{2} \left( C^{(S)}_{\ell H} Y^{}_l Y^\dagger_l  \right)^{\alpha\beta} + \left( 2\lambda - \frac{3}{2} g^2_2 \right) \left( C^{}_{\ell e HD} Y^\dagger_l \right)^{\alpha\beta} 
	\nonumber
	\\
	&& + \left( C^{}_{\ell e HD} Y^\dagger_l Y^{}_l Y^\dagger_l \right)^{\alpha\beta} - \frac{3}{4} g^2_2  \left(  g^2_2 - 4 \lambda \right) C^{(S)\alpha\beta}_{\ell HD1} + \lambda \left( C^{(S)}_{\ell HD1} Y^{}_l  Y^\dagger_l \right)^{\alpha\beta} 
	\nonumber
	\\
	&& - \left( C^{(S)}_{\ell HD1} Y^{}_l  Y^\dagger_l Y^{}_l Y^\dagger_l  \right)^{\alpha\beta} - \frac{3}{8} \left( g^4_1 + 2 g^2_1 g^2_2 + 3g^4_2 - 4g^2_2 \lambda \right) C^{(S)\alpha\beta}_{\ell HD2} - \frac{1}{2} \lambda \left( C^{(S)}_{\ell HD2} Y^{}_l  Y^\dagger_l \right)^{\alpha\beta} 
	\nonumber
	\\
	&& - \left( C^{(S)}_{\ell HD2} Y^{}_l  Y^\dagger_l Y^{}_l Y^\dagger_l  \right)^{\alpha\beta} - 3 g^3_2 C^{\alpha\beta}_{\ell HW} - 6 g^{}_2 \left( C^{}_{\ell HW} Y^{}_l Y^\dagger_l \right)^{\alpha\beta} - 3 C^{(S)\gamma\lambda\alpha\beta}_{\overline{e}\ell\ell\ell H} \left[ \lambda \left( Y^{}_l \right)^{}_{\lambda\gamma} \right. 
	\nonumber
	\\
	&& - \left. \left( Y^{}_l Y^\dagger_l Y^{}_l \right)^{}_{\lambda\gamma}  \right] - 2C^{(M)\gamma\lambda\alpha\beta}_{\overline{e}\ell\ell\ell H}   \left[  \lambda  \left( Y^{}_l \right)^{}_{\lambda\gamma} - \left( Y^{}_l Y^\dagger_l Y^{}_l \right)^{}_{\lambda\gamma} \right]  - 3 C^{\gamma\alpha\lambda\beta}_{\overline{d}\ell q\ell H1} \left[ \lambda  \left( Y^{}_{\rm d} \right)^{}_{\lambda\gamma} \right.
	\nonumber
	\\
	&& - \left. \left( Y^{}_{\rm d} Y^\dagger_{\rm d} Y^{}_{\rm d} \right)^{}_{\lambda\gamma}  \right] + 6 C^{\gamma\lambda\alpha\beta}_{\overline{q}u\ell\ell H} \left[ \lambda \left( Y^\dagger_{\rm u} \right)^{}_{\lambda\gamma} - \left( Y^\dagger_{\rm u} Y^{}_{\rm u} Y^\dagger_{\rm u} \right)^{}_{\lambda\gamma} \right] + \alpha \leftrightarrow \beta \;.
\end{eqnarray}

\noindent$\bullet~\bm{\psi^2H^3D}$

\begin{eqnarray}
	\befun^{\alpha\beta}_{\ell e HD} &=& - \frac{3}{2} \left( 3g^2_1 - 4\lambda - 2T \right) C^{\alpha\beta}_{\ell eHD} + \left( Y^{\rm T}_l C^{}_{\ell eHD} Y^\dagger_l \right)^{\beta\alpha} + 4 \left( C^{}_{\ell e HD} Y^\dagger_l Y^{}_l \right)^{\alpha\beta} + \frac{1}{2} \left( C^{\rm T}_{\ell e HD} Y^{}_l Y^\dagger_l \right)^{\beta\alpha} 
	\nonumber
	\\
	&& + \left( 3g^2_1 - g^2_2 \right) \left( C^{(S)}_{\ell HD1} Y^{}_l \right)^{\alpha\beta} - 2\left( C^{(S)}_{\ell HD1} Y^{}_l Y^\dagger_l Y^{}_l \right)^{\alpha\beta} + \frac{1}{8} \left( 7g^2_1 - 17 g^2_2 - 8\lambda \right) \left( C^{(S)}_{\ell HD2} Y^{}_l \right)^{\alpha\beta} 
	\nonumber
	\\
	&& - 2\left( C^{(S)}_{\ell HD2} Y^{}_l Y^\dagger_l Y^{}_l \right)^{\alpha\beta}  - \frac{1}{2}  \left( Y^{\rm T}_l C^{(S)}_{\ell HD2} Y^{}_l Y^\dagger_l \right)^{\beta\alpha} - 6 C^{\gamma\alpha\lambda\beta}_{\overline{d}\ell ue H} \left( Y^\dagger_{\rm u} Y^{}_{\rm d} \right)^{}_{\lambda\gamma} \;.
\end{eqnarray}

\noindent$\bullet~\bm{\psi^2H^2D^2}$

\begin{eqnarray}
	\befun^{(S)\alpha\beta}_{\ell HD1} &=& -\frac{1}{4} \left( 3g^2_1 - 11g^2_2 - 4T \right) C^{(S)\alpha\beta}_{\ell HD1} + \frac{7}{2} \left( C^{(S)}_{\ell HD1} Y^{}_l Y^\dagger_l \right)^{\alpha\beta} - \frac{1}{8} \left( 11 g^2_1 + 11 g^2_2 + 8\lambda \right) C^{(S)\alpha\beta}_{\ell HD2} 
	\nonumber
	\\
	&& + 6 C^{(S)\gamma\lambda\alpha\beta}_{\overline{d}u\ell\ell D} \left( Y^\dagger_{\rm u} Y^{}_{\rm d } \right)^{}_{\lambda\gamma} + \alpha \leftrightarrow \beta \;,
	\nonumber
	\\
	\befun^{(S)\alpha\beta}_{\ell HD2} &=& -4g^2_2 C^{(S)\alpha\beta}_{\ell HD1} - 4 \left( C^{(S)}_{\ell HD1} Y^{}_l Y^\dagger_l \right)^{\alpha\beta} + \frac{1}{2} \left( 4 g^2_1 + g^2_2 + 4\lambda + 2T \right) C^{(S)\alpha\beta}_{\ell HD2} - \frac{3}{2} \left( C^{(S)}_{\ell HD2} Y^{}_l Y^\dagger_l \right)^{\alpha\beta} 
	\nonumber
	\\
	&& + \alpha \leftrightarrow \beta \;.
\end{eqnarray}

\noindent$\bullet~\bm{\psi^2H^2X}$
\begin{eqnarray}
	\befun^{\alpha\beta}_{\ell HW} &=& \frac{1}{2} g^3_2 C^{(S)\alpha\beta}_{\ell HD1} - \frac{1}{4} g^{}_2  \left( C^{(S)}_{\ell HD1} Y^{}_l Y^\dagger_l \right)^{\beta\alpha} + \frac{1}{2} g^{}_2  \left( C^{(S)}_{\ell HD1} Y^{}_l Y^\dagger_l \right)^{\alpha\beta} + \frac{5}{8} g^3_2 C^{(S)\alpha\beta}_{\ell HD2} 
	\nonumber
	\\
	&& + \frac{3}{4} g^{}_2 \left( C^{(S)}_{\ell HD2} Y^{}_l Y^\dagger_l \right)^{\alpha\beta} + \frac{1}{8} g^{}_2 \left( C^{(S)}_{\ell HD2} Y^{}_l Y^\dagger_l \right)^{\beta\alpha} - \frac{1}{2} \left( 4g^2_1 - 9g^2_2 - 8\lambda - 4T  \right) C^{\alpha\beta}_{\ell HW} 
	\nonumber
	\\
	&& + \frac{7}{2} g^2_2 C^{\beta\alpha}_{\ell HW} + \frac{9}{2} \left( C^{}_{\ell HW} Y^{}_l Y^\dagger_l \right)^{\alpha\beta}  + 2\left( C^{}_{\ell HW} Y^{}_l Y^\dagger_l \right)^{\beta\alpha} - \frac{3}{2} \left( C^{\rm T}_{\ell HW} Y^{}_l Y^\dagger_l \right)^{\beta\alpha} 
	\nonumber
	\\
	&& + 3 g^{}_1 g^{}_2 C^{(A)\alpha\beta}_{\ell HB} - \frac{1}{4} g^{}_2 \left( 3 C^{(S)\gamma\lambda\alpha\beta}_{\overline{e}\ell\ell\ell H} + C^{(A)\gamma\lambda\alpha\beta}_{\overline{e}\ell\ell\ell H} - 2C^{(M)\gamma\lambda\alpha\beta}_{\overline{e}\ell\ell\ell H} \right) \left( Y^{}_l \right)^{}_{\lambda\gamma} 
	\nonumber
	\\
	&& - \frac{3}{4} g^{}_2 C^{\gamma\beta\lambda\alpha}_{\overline{d}\ell q \ell H1} \left( Y^{}_{\rm d} \right)^{}_{\lambda\gamma} - \frac{3}{4} g^{}_2 \left( C^{\gamma\alpha\lambda\beta}_{\overline{d}\ell q \ell H2} + C^{\gamma\beta\lambda\alpha}_{\overline{d}\ell q \ell H2} \right) \left( Y^{}_{\rm d} \right)^{}_{\lambda\gamma} \;,
	\nonumber
	\\
	\befun^{(A)\alpha\beta}_{\ell HB} &=& - \frac{1}{8} g^{}_1 \left( C^{(S)}_{\ell HD1} Y^{}_l Y^\dagger_l \right)^{\alpha\beta}  - \frac{3}{16} g^{}_1 \left( C^{(S)}_{\ell HD2} Y^{}_l Y^\dagger_l \right)^{\alpha\beta} + 3 g^{}_1 g^{}_2 C^{\alpha\beta}_{\ell HW} - \frac{3}{2} \left( C^{(A)}_{\ell HB} Y^{}_l Y^\dagger_l \right)^{\alpha\beta} 
	\nonumber
	\\
	&& + \frac{1}{12} \left( 47g^2_1 - 30g^2_2 + 24 \lambda + 12T \right) C^{(A)\alpha\beta}_{\ell HB} + \frac{3}{8} g^{}_1 \left( C^{(A)\gamma\lambda\alpha\beta}_{\overline{e}\ell\ell\ell H} - 2 C^{(M)\gamma\lambda\alpha\beta}_{\overline{e}\ell\ell\ell H} \right) \left( Y^{}_l \right)^{}_{\lambda\gamma}
	\nonumber
	\\
	&& - \frac{1}{8} g^{}_1 C^{\gamma\alpha\lambda\beta}_{\overline{d}\ell q\ell H1}\left( Y^{}_{\rm d} \right)^{}_{\lambda\gamma} - \alpha \leftrightarrow \beta \;.
\end{eqnarray}

\noindent$\bullet~\bm{\psi^4H}$


\noindent$\bullet~\bm{\psi^4D}$
\begin{eqnarray}\label{eq:psi4D}
	\befun^{(S)\alpha\beta\gamma\lambda}_{\overline{e}dddD} &=&  -\frac{2}{3} \left( g^2_1 - 6g^2_3 \right) C^{(S)\alpha\beta\gamma\lambda}_{\overline{e}dddD}  +  C^{(S)\rho\beta\gamma\lambda}_{\overline{e}dddD}  \left(Y^\dagger_l Y^{}_l \right)^{}_{\alpha\rho} +  C^{(S)\alpha\rho\gamma\lambda}_{\overline{e}dddD} \left( Y^\dagger_{\rm d} Y^{}_{\rm d} \right)^{}_{\rho\beta} 
	\nonumber
	\\
	&& +  C^{(S)\alpha\beta\rho\lambda}_{\overline{e}dddD} \left( Y^\dagger_{\rm d} Y^{}_{\rm d} \right)^{}_{\rho\gamma} + C^{(S)\alpha\beta\gamma\rho}_{\overline{e}dddD} \left( Y^\dagger_{\rm d} Y^{}_{\rm d} \right)^{}_{\rho\lambda}
	\nonumber
	\\
	&& - \frac{2}{3} \left( Y^\dagger_l \right)^{}_{\alpha\rho}  \left[ C^{(S)\rho\sigma\gamma\lambda}_{\overline{\ell}qddD} \left( Y^{}_{\rm d} \right)^{}_{\sigma\beta} + C^{(S)\rho\sigma\beta\lambda}_{\overline{\ell}qddD} \left( Y^{}_{\rm d} \right)^{}_{\sigma\gamma} + C^{(S)\rho\sigma\beta\gamma}_{\overline{\ell}qddD} \left( Y^{}_{\rm d} \right)^{}_{\sigma\lambda}  \right] \;,
	\nonumber
	\\
	\befun^{(S)\alpha\beta\gamma\lambda}_{\overline{d}u\ell\ell D} &=& \left( 2 C^{(S)\gamma\lambda}_{\ell HD1} + C^{(S)\gamma\lambda}_{\ell HD2} \right) \left( Y^\dagger_{\rm d} Y^{}_{\rm u} \right)^{}_{\alpha\beta} + \frac{1}{6} \left( g^2_1 + 9g^2_2 \right) C^{(S)\alpha\beta\gamma\lambda}_{\overline{d}u\ell\ell D} +  C^{(S)\rho\beta\gamma\lambda}_{\overline{d}u\ell\ell D} \left( Y^\dagger_{\rm d} Y^{}_{\rm d} \right)^{}_{\alpha\rho} 
	\nonumber
	\\
	&& + C^{(S)\alpha\rho\gamma\lambda}_{\overline{d}u\ell\ell D} \left( Y^\dagger_{\rm u} Y^{}_{\rm u} \right)^{}_{\rho\beta} + \frac{1}{2} C^{(S)\alpha\beta\rho\lambda}_{\overline{d}u\ell\ell D} \left( Y^{}_l Y^\dagger_l \right)^{}_{\rho\gamma} + \frac{1}{2} C^{(S)\alpha\beta\gamma\rho}_{\overline{d}u\ell\ell D} \left( Y^{}_l Y^\dagger_l \right)^{}_{\rho\lambda} \;,
	\nonumber
	\\
	\befun^{(S)\alpha\beta\gamma\lambda}_{\overline{\ell}qddD} &=& - 3 C^{(S)\rho\sigma\gamma\lambda}_{\overline{e}dddD} \left( Y^{}_l \right)^{}_{\alpha\rho} \left( Y^\dagger_{\rm d} \right)^{}_{\sigma\beta} + \frac{4}{9} \left( g^2_1 + 3g^2_3 \right) C^{(S)\alpha\beta\gamma\lambda}_{\overline{\ell}qddD} - C^{(S)\alpha\rho\sigma\lambda}_{\overline{\ell}qddD} \left( Y^\dagger_{\rm d} \right)^{}_{\sigma\beta} \left( Y^{}_{\rm d} \right)^{}_{\rho\gamma}
	\nonumber
	\\
	&&  - C^{(S)\alpha\rho\gamma\sigma}_{\overline{\ell}qddD} \left( Y^\dagger_{\rm d} \right)^{}_{\sigma\beta} \left( Y^{}_{\rm d} \right)^{}_{\rho\lambda} + \frac{1}{2} C^{(S)\rho\beta\gamma\lambda}_{\overline{\ell}qddD} \left( Y^{}_l Y^\dagger_l \right)^{}_{\alpha\rho} + \frac{1}{2} C^{(S)\alpha\rho\gamma\lambda}_{\overline{\ell}qddD} \left[ \left( Y^{}_{\rm d} Y^\dagger_{\rm d} \right)^{}_{\rho\beta} +  \left( Y^{}_{\rm u} Y^\dagger_{\rm u} \right)^{}_{\rho\beta} \right]
	\nonumber
	\\
	&& +  C^{(S)\alpha\beta\rho\lambda}_{\overline{\ell}qddD}  \left( Y^{}_{\rm d} Y^\dagger_{\rm d} \right)^{}_{\rho\gamma} +  C^{(S)\alpha\beta\gamma\rho}_{\overline{\ell}qddD}  \left( Y^{}_{\rm d} Y^\dagger_{\rm d} \right)^{}_{\rho\lambda} \;.
\end{eqnarray}

\section{Analysis and Partial Check}\label{sec:analysis}

\subsection{Structure of the Anomalous Dimension Matrix}

\begin{table}
	\centering
	\renewcommand\arraystretch{1.9}
	\resizebox{\textwidth}{!}{
		\begin{tabular}{c|cc|c|cc|ccc|c|cc|c}
			\hline\hline
			\diagbox{\makecell[c]{ $C^{}_i$\\ $(w_i,\overline{w}_i)$}}{$\gamma^{}_{ij}$}{\makecell[c]{ $C^{}_j$\\ $(w_j,\overline{w}_j)$} } & \makecell[c]{ $C^{(S)}_{\ell HD1}$ \\ $(3,5)$} & \makecell[c]{ $C^{(S)}_{\ell HD2}$ \\ $(3,5)$}  & \makecell[c]{ $C^{(S)}_{\overline{d}u\ell\ell D}$ \\ $(3,5)$} & \makecell[c]{ $C^{(A)}_{\ell HB}$  \\ $(3,7)$} & \makecell[c]{$C^{}_{\ell HW}$ \\ $(3,7)$} & \makecell[c]{ $C^{(S,A,M)}_{\overline{e}\ell\ell\ell H}$  \\ $(3,7)$}  & \makecell[c]{ $C^{}_{\overline{d}\ell q\ell H1}$ \\ $(3,7)$} & \makecell[c]{ $C^{}_{\overline{d}\ell q\ell H2}$ \\ $(3,7)$} & \makecell[c]{ $C^{}_{\ell eHD}$ \\ $(5,5)$}  & \makecell[c]{ $C^{}_{\overline{d}\ell ueH}$ \\ $(5,5)$} & \makecell[c]{ $C^{}_{\overline{q}u\ell\ell H}$ \\ $(5,5)$} & \makecell[c]{ $C^{(S)}_{\ell H}$\\ $(5,7)$} \\
			\hline
			\makecell[c]{ $C^{(S)}_{\ell HD1}$ \\ $(3,5)$}  & $g^2,y^2$ & $g^2,\lambda$ & $y^2$ & \cellcolor{gray!30}{0}  & \cellcolor{gray!30}{0}  & \cellcolor{gray!30}{0}  & \cellcolor{gray!30}{0}  & \cellcolor{gray!30}{0} & \cellcolor{gray!30}{0}  & \cellcolor{gray!30}{0}  & \cellcolor{gray!30}{0} & \cellcolor{gray!30}{0}  \\
			\makecell[c]{ $C^{(S)}_{\ell HD2}$ \\ $(3,5)$}  & $g^2,y^2$ & $g^2,y^2, \lambda$ & 0  & \cellcolor{gray!30}{0}  & \cellcolor{gray!30}{0}  & \cellcolor{gray!30}{0}  & \cellcolor{gray!30}{0}  & \cellcolor{gray!30}{0} & \cellcolor{gray!30}{0}  & \cellcolor{gray!30}{0} & \cellcolor{gray!30}{0}  & \cellcolor{gray!30}{0} \\
			\hline
			\makecell[c]{ $C^{(S)}_{\overline{d}u\ell\ell D}$ \\ $(3,5)$} &  $y^2$ & $y^2$ &  $g^2, y^2$ &  \cellcolor{gray!30}{0} &  \cellcolor{gray!30}{0} & \cellcolor{gray!30}{0} &  \cellcolor{gray!30}{0} &  \cellcolor{gray!30}{0} &  \cellcolor{gray!30}{0}  &  \cellcolor{gray!30}{0} &  \cellcolor{gray!30}{0} &  \cellcolor{gray!30}{0} \\
			\hline
			\makecell[c]{ $C^{(A)}_{\ell HB}$  \\ $(3,7)$} & $g y^2$ & $gy^2$  & 0 & $g^2, y^2, \lambda$ & $g^2$  & $gy$ & $gy$ & 0 & \cellcolor{gray!30}{0} & \cellcolor{gray!30}{0}  & \cellcolor{gray!30}{0} & \cellcolor{gray!30}{0} \\
			\makecell[c]{$C^{}_{\ell HW}$ \\ $(3,7)$} & $g^3, g y^2$ & $g^3, gy^2$ & 0 & $g^2$ & $g^2, y^2, \lambda$  & $gy$ & $gy$ & $gy$  & \cellcolor{gray!30}{0} & \cellcolor{gray!30}{0}  & \cellcolor{gray!30}{0}  & \cellcolor{gray!30}{0} \\
			\hline
			\makecell[c]{ $C^{(S,A,M)}_{\overline{e}\ell\ell\ell H}$  \\ $(3,7)$} & $g^2y, y^3$ & $g^2y, y^3$ & 0 & $gy$ & $gy$ &  $g^2, y^2$ & $y^2$ & $y^2$ & \cellcolor{gray!30}{0} & \cellcolor{gray!30}{0} & \cellcolor{gray!70}{$\overline{y}^2$} & \cellcolor{gray!30}{0} \\
			\makecell[c]{ $C^{}_{\overline{d}\ell q\ell H1}$ \\ $(3,7)$}  & $g^2y, y^3$ & $g^2y, y^3$ & $g^2y,y^3$ & $gy$ & $gy$ & $ y^2$ & $g^2, y^2$ & $g^2, y^2$ & \cellcolor{gray!30}{0} & \cellcolor{gray!70}{$\overline{y}^2$} & \cellcolor{gray!70}{$\overline{y}^2$} & \cellcolor{gray!30}{0} \\
			\makecell[c]{ $C^{}_{\overline{d}\ell q\ell H2}$ \\ $(3,7)$} & $g^2y, y^3$ & $g^2y, y^3$ & $g^2y,y^3$ & $gy$ & $gy$ & $ y^2$ & $g^2, y^2$ & $g^2, y^2$ & \cellcolor{gray!30}{0} & \cellcolor{gray!70}{$\overline{y}^2$} & \cellcolor{gray!70}{$\overline{y}^2$} & \cellcolor{gray!30}{0} \\
			\hline
			\makecell[c]{ $C^{}_{\ell eHD}$ \\ $(5,5)$}  &$g^2y, y^3$ & $g^2y, \lambda y, y^3$ & 0 &  \cellcolor{gray!30}{0}  &  \cellcolor{gray!30}{0} & \cellcolor{gray!30}{0}  &  \cellcolor{gray!30}{0} &  \cellcolor{gray!30}{0}  & $g^2,y^2,\lambda$  & $y^2$ & 0 &    \cellcolor{gray!30}{0} \\
			\hline
			\makecell[c]{ $C^{}_{\overline{d}\ell ueH}$ \\ $(5,5)$}  & $y^3$ & $y^3$ & $g^2y, y^3$ & \cellcolor{gray!30}{0}  & \cellcolor{gray!30}{0} & \cellcolor{gray!30}{0}  & \cellcolor{gray!70}{$y^2$} & \cellcolor{gray!70}{$y^2$}  & $y^2$ & $g^2,y^2$ & $y^2$  & \cellcolor{gray!30}{0}  \\
			\makecell[c]{ $C^{}_{\overline{q}u\ell\ell H}$ \\ $(5,5)$} & $g^2y, y^3$ & $g^2y, y^3$ & $g^2y, y^3$ & \cellcolor{gray!30}{0}  & \cellcolor{gray!30}{0} & \cellcolor{gray!70}{$y^2$}  & \cellcolor{gray!70}{$y^2$} & \cellcolor{gray!70}{$y^2$}  & 0 & $y^2$ & $g^2, y^2$ & \cellcolor{gray!30}{0}  \\
			\hline
			\makecell[c]{ $C^{(S)}_{\ell H}$\\ $(5,7)$} & \makecell[c]{ $g^4,\lambda g^2, $ \\ $\lambda y^2, y^4$} & \makecell[c]{ $g^4,\lambda g^2, $ \\ $\lambda y^2, y^4$} & 0 & 0 & $g^3,gy^2$ & $\lambda y, y^3$ & $\lambda y, y^3$ & 0 & $\lambda y, g^2 y , y^3$ & 0 & $\lambda y, y^3$ & $g^2,y^2,\lambda$ \\
			\hline\hline
		\end{tabular}
	}
	\\[0.4cm]
	\resizebox{0.53\textwidth}{!}{
		\begin{tabular}{c|cc|cc|cc}
			\hline\hline
			\diagbox{\makecell[c]{ $C^{}_i$\\ $(w_i,\overline{w}_i)$} }{$\gamma^{}_{ij}$}{\makecell[c]{ $C^{}_i$\\ $(w_j,\overline{w}_j)$} } & \makecell[c]{ $C^{(S)}_{\overline{e}dddD}$ \\ $(5,3)$ } & \makecell[c]{ $C^{(S)}_{\overline{\ell}qddD}$ \\ $(5,3)$ } & \makecell[c]{ $C^{(A)}_{\overline{e}qdd\widetilde{H}}$ \\ $(5,5)$ } & \makecell[c]{ $C^{}_{\overline{\ell} dqq\widetilde{H}}$ \\ $(5,5)$ } & \makecell[c]{ $C^{}_{\overline{\ell} dud\widetilde{H}}$ \\ $(7,3)$ } &  \makecell[c]{ $C^{(M)}_{\overline{\ell}dddH}$ \\ $(7,3)$ } \\
			\hline
			\makecell[c]{ $C^{(S)}_{\overline{e}dddD}$ \\ $(5,3)$ } & $g^2,y^2$ & $y^2$ & \cellcolor{gray!30}{0} & \cellcolor{gray!30}{0} & \cellcolor{gray!30}{0} & \cellcolor{gray!30}{0} \\
			\makecell[c]{ $C^{(S)}_{\overline{\ell}qddD}$ \\ $(5,3)$ } & $y^2$ & $g^2,y^2$ & \cellcolor{gray!30}{0} & \cellcolor{gray!30}{0} & \cellcolor{gray!30}{0} & \cellcolor{gray!30}{0} \\
			\hline
			\makecell[c]{ $C^{(A)}_{\overline{e}qdd\widetilde{H}}$ \\ $(5,5)$ } & $g^2y, y^3$ & $y^3$ & $g^2,y^2$ & $y^2$ & \cellcolor{gray!70}{$\overline{y}^2$}  & \cellcolor{gray!30}{0} \\
			\makecell[c]{ $C^{}_{\overline{\ell} dqq\widetilde{H}}$ \\ $(5,5)$ } & $ y^3$ & $g^2y, y^3$ & $y^2$ & $g^2, y^2$ & \cellcolor{gray!70}{$\overline{y}^2$}  & \cellcolor{gray!30}{0} \\
			\hline
			\makecell[c]{ $C^{}_{\overline{\ell} dud\widetilde{H}}$ \\ $(7,3)$ } & $y^3$ & $g^2y,y^3$ & \cellcolor{gray!70}{$y^2$} & \cellcolor{gray!70}{$y^2$} & $g^2,y^2$ & $y^2$ \\
			\makecell[c]{ $C^{(M)}_{\overline{\ell}dddH}$ \\ $(7,3)$ } & $y^3$ & $g^2y,y^3$ & \cellcolor{gray!30}{0} & \cellcolor{gray!30}{0} & $y^2$ & $g^2,y^2$ \\
			\hline\hline
		\end{tabular}
	}
	\caption{The structure and perturbative power counting of the one-loop anomalous dimension matrix $\gamma^{}_{ij}$ for Wilson coefficients of dim-7 baryon-number-conserving (the upper one) and baryon-number-violating (the lower one) operators. $w$ and $\overline{w}$ are holomorphic and antiholomorphic weights of operators and $g$, $y(\overline{y})$, $\lambda$ respectively denote gauge couplings, Yukawa couplings and Higgs quartic coupling with $g$, $y(\overline{y})$ and $\sqrt{\lambda}$ being at the same order for  perturbative power counting.}
	\label{tab:adm}
\end{table}

With the help of one-loop RGEs of dim-7 operators in Eqs.~\eqref{eq:psi2H4}-\eqref{eq:psi4D}, one may extract the structure and perturbative power counting of the one-loop anomalous dimension matrix $\gamma^{}_{ij}$ for Wilson coefficients of dim-7 operators, which are summarized in Table~\ref{tab:adm}. The upper and lower tables are for BNC and BNV operators, respectively, which do not mix with each other. The holomorphic and antiholomorphic weights $(w,\overline{w})$ associated with operators are also attached to each Wilson coefficient. In Table~\ref{tab:adm}, $g$, $y(\overline{y})$ and $\lambda$ denote gauge couplings, (conjugate) Yukawa couplings and Higgs quartic coupling, and they characterize the perturbative power counting order of $\gamma^{}_{ij}$ with $g$, $y(\overline{y})$ and $\sqrt{\lambda}$ being at the same order for perturbative power counting. We only explicitly distinguish between Yukawa and conjugate Yukawa couplings for the entries in darker grey cells of Table~\ref{tab:adm}. The structure shown in Table~\ref{tab:adm} can be well understood by taking advantage of a non-renormalization theorem~\cite{Cheung:2015aba}. In accordance with the nonrenormalization theorem, if an operator $\Op^{}_j$ can renormalize an  operator $\Op^{}_i$ at the one-loop level, their holomorphic and antiholomorphic weights should satisfy two inequalities, i.e., $w^{}_i \geq w^{}_j$ and $\overline{w}_i \geq \overline{w}^{}_j$ but Yukawa couplings of a  nonholomorphic form can violate these inequalities. In other words, entries of the one-loop anomalous dimension matrix should be zero, i.e., $\gamma^{}_{ij} = 0$ if $w^{}_i < w^{}_j$ or $\overline{w}^{}_i < \overline{w}^{}_j$ holds and nonholomorphic Yukawa couplings are absent. This can perfectly explain the zero entries in lighter grey cells of Table~\ref{tab:adm}, where $w^{}_i < w^{}_j$ or $\overline{w}^{}_i < \overline{w}^{}_j$. The entries in darker grey cells with $w^{}_i < w^{}_j$ or $\overline{w}^{}_i < \overline{w}^{}_j$ are non-vanishing but have Yukawa couplings of nonholomorphic forms, i.e., $y^2$ and $\overline{y}^2$. In the SMEFT, those nonholomorphic Yukawa couplings, more specifically, $Y^{}_{\rm u} Y^{}_{{\rm d},l}$ and $Y^\dagger_{\rm u} Y^\dagger_{{\rm d},l}$ come from exchanging a Higgs doublet via Yukawa interactions $\overline{Q^{}_{\rm L}} Y^{}_{\rm u} \widetilde{H} U^{}_{\rm R}$ and $\overline{Q^{}_{\rm L}} Y^{}_{\rm d} H D^{}_{\rm R}$ or $\overline{\ell^{}_{\rm L}} Y^{}_l H E^{}_{\rm R}$ and via their conjugates as well. This exactly corresponds to one of the exceptional four-point tree amplitudes and violates the aforementioned inequalities by two units as pointed out in Ref.~\cite{Cheung:2015aba}. Therefore, the structure given in Table~\ref{tab:adm} is fully consistent with the non-renormalization theorem in Ref.~\cite{Cheung:2015aba}. As can be seen from Table~\ref{tab:adm}, for the BNC operators, $\Op^{(S)}_{\ell H}$ has the largest holomorphic and nonholomorphic weights and hence only renormalizes itself. On the contrary, $\Op^{(S)}_{\ell HD1}$, $\Op^{(S)}_{\ell HD2}$ and $\Op^{(S)}_{\overline{d}u\ell\ell D}$ have the smallest holomorphic and nonholomorphic weights, thus they have the potential for renormalizing all BNC operators according to the non-renormalization theorem. This is indeed the case for $\Op^{(S)}_{\ell HD1}$, $\Op^{(S)}_{\ell HD2}$, but not for $\Op^{(S)}_{\overline{d}u\ell\ell D}$ since some one-loop diagrams with an insertion of $\Op^{(S)}_{\overline{d}u\ell\ell D}$ for other operators are absent. Similar properties can be found from Table~\ref{tab:adm} for the BNV operators.

On the other hand, the perturbative order of non-zero entries of $\gamma^{}_{ij}$ can be accounted for by the naive dimensional analysis (NDA)~\cite{Jenkins:2013sda} or power counting rules~\cite{Liao:2017amb}. The latter one applied to dim-7 operators has been studied in detail in Refs.~\cite{Liao:2017amb,Liao:2019tep}, which produces a generic perturbative order for each entry matching well those in Table~\ref{tab:adm}. Here, we make use of the NDA to briefly analyse the perturbative order of $\gamma^{}_{ij}$ for dim-7 operators. The NDA-weights of the six dim-7 operator classes $\{ \psi^2 H^4, \psi^2 H^3 D, \psi^4 H, \psi^2 H^2 D^2, g\psi^2 H^2 X, \psi^4 D \}$ are $\omega_i =\{ 2, 3/2, 3/2, 1, 1, 1 \}$ (for $i=1,\dots, 6$) with $\psi^2 H^2 X$ operators normalized by a factor $g=g^{}_1, g^{}_2$ for $X= B, W$. Then the perturbative weight $N^{}_{ij}$ of $\gamma^{}_{ij}$ is determined by $N^{}_{ij} = 1 + \omega^{}_i - \omega^{}_j$ at the one-loop level, whose entries are found to be
\begin{eqnarray}\label{eq:nda}
	\begin{blockarray}{rccc}
		& \left\{ \psi^2 H^4 \right\} &  \left\{ \psi^2 H^3 D, \psi^4 H \right\} & \left\{ \psi^2 H^2 D, g\psi^2 H^2 X, \psi^4 D \right\}  \\[5pt]
		\begin{block}{r(ccc)}
			\left\{ \psi^2 H^4 \right\} \phantom{a} & 1 & \displaystyle \frac{3}{2}  & 2 \\[5pt]
			\left\{  \psi^2 H^3 D, \psi^4 H \right\} \phantom{a} &  \displaystyle\frac{1}{2}  &  1 & \displaystyle\frac{3}{2}  \\[5pt]
			\left\{ \psi^2 H^2 D, g\psi^2 H^2 X, \psi^4 D \right\} \phantom{a} &0 & \displaystyle\frac{1}{2} & 1 \\
		\end{block}
	\end{blockarray} \;,
\end{eqnarray}
where classes with the same NDA-weight are collected. The perturbative weight $N$ is defined as $N\equiv n^{}_g + n^{}_y + n^{}_\lambda$ in the SMEFT with $n^{}_{g,y,\lambda}$ being respectively the powers of $g^2/\left( 16\pi^2 \right)$, $y^2/\left( 16\pi^2 \right)$ and $\lambda/\left( 16\pi^2 \right)$ with NDA normalization. From this definition, one easily understands why $N$ is called the perturbative order, i.e., $N$ is the total power of the perturbative factor $1/(16\pi^2)$. With the help of Eq.~\eqref{eq:nda}, one can figure out the generic perturbative order of $\gamma^{}_{ij}$ which is coincident with that in Table~\ref{tab:adm} after taking into account normalization factor $g$ for $\psi^2 H^2X$ operators. For example, the generic perturbative order is $  \left( {\mathbbm{g}}^2 \right)^{1/2} \times \mathbbm{g} = \mathbbm{g}^2 $ for $\psi^2H^2X$-$\psi^4 H$ and $ \left( {\mathbbm{g}}^2 \right)^{2} / \mathbbm{g} = \mathbbm{g}^3 $ for $\psi^2 H^4$-$\psi^2 H^2 X$ with $\mathbbm{g}$ of the same order as $g,y,\sqrt{\lambda}$.

\subsection{Partial Check on Results}
We find there is an obvious discrepancy between our results for the anomalous dimensions of $C^{(A)}_{\ell  HB}$ and $C^{}_{\ell HW}$ and those in Ref.~\cite{Liao:2019tep}. As can be easily seen from Table~\ref{tab:adm} in this work and a similar table in Ref.~\cite{Liao:2019tep}, contributions from $C^{}_{\ell eHD}$ to $C^{(A)}_{\ell  HB}$ and $C^{}_{\ell HW}$ are vanishing in this work but nonvanishing, i.e., $g\overline{y}$ (after taking into account different notations) in Ref.~\cite{Liao:2019tep}. Actually, such a single Yukawa-coupling contribution can not result from the unique exceptional amplitude $\mathcal{A} \left( \psi^+\psi^+\psi^+\psi^+ \right)$ in the SM(EFT)~\cite{Cheung:2015aba,Craig:2019wmo} and hence should not appear~\footnote{We have made contact with the authors of Ref.~\cite{Liao:2019tep} for this discrepancy. After carefully checking on their results, they found that a relevant topology was overlooked in their calculation and agreed on our results.}. In order to check this discrepancy, we recalculate contributions from the operator $\Op^{}_{\ell eHD}$ to the counterterms for $\Op^{(A)}_{\ell HB}$ and $\Op^{}_{\ell HW}$ by means of {\sf FeynRules}~\cite{Christensen:2008py,Alloul:2013bka}, {\sf FeynArts}~\cite{Hahn:2000kx}, and {\sf FeynCalc}~\cite{Shtabovenko:2016sxi,Shtabovenko:2020gxv} with the general $R^{}_{\xi}$ gauge for quantum gauge fields. Additionally, we reproduce all contributions from the operator $\Op^{(S)}_{\ell HD1}$ in the same way as a crosscheck for our results obtained by {\sf Matchmakereft} with the Feynman-'t Hooft gauge.

\subsubsection{Contributions to $C^{(A)}_{\ell HB}$ and $C^{}_{\ell HW}$ from $C^{}_{\ell eHD}$} 

\begin{figure}
	\centering
	\includegraphics[width=\linewidth]{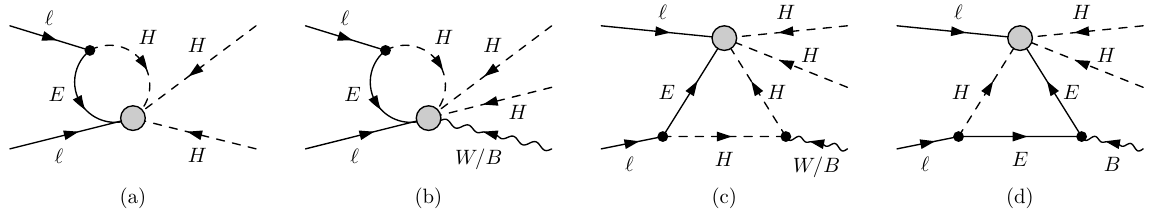}
	\vspace{-0.8cm}
	\caption{The relevant Feynman diagrams with an insertion of the operator $\Op^{}_{\ell eHD}$ for calculating counterterms of $\psi^2 H^2 D^2$ and $\psi^2H^2 X$ operators in the Green's basis.}
	\label{fig:fd}
\end{figure}

As can be seen from the reduction relations for dim-7 operators in Ref.~\cite{Zhang:2023kvw}, $\psi^2H^2D^2$-class operators in the Green's basis can be reduced to $\Op^{(A)}_{\ell HB}$ and $\Op^{}_{\ell HW}$. Therefore, we have to work out counterterms not only for $\Op^{(A)}_{\ell HB}$ and $\Op^{}_{\ell HW}$ themselves but also for those in the $\psi^2H^2D^2$ class in the Green's basis. All relevant Feynman diagrams with an insertion of $\Op^{}_{\ell eHD}$ are shown in Fig.~\ref{fig:fd}, where the first one is used to reach the counterterms for $\psi^2H^2D^2$ operators and the rest for $\Op^{(A)}_{\ell HB}$ or $\Op^{}_{\ell HW}$. After figuring out UV divergences of all diagrams in Fig.~\ref{fig:fd} by means of {\sf FeynRules}, {\sf FeynArts}, and {\sf FeynCalc}, one acquires the following counterterms to cancel out all relevant UV divergences in the Green's basis:
\begin{eqnarray}\label{eq:counterterm-lehd}
	\delta G^{\alpha\beta}_{\ell HD1} &=& \delta G^{(S)\alpha\beta}_{\ell HD3} = 0 \;,
	\nonumber
	\\
	\delta G^{\alpha\beta}_{\ell HD2} &=& - \frac{3}{2} \left( C^{}_{\ell eHD} Y^\dagger_l \right)^{\alpha\beta} \;,
	\nonumber
	\\
	\delta G^{(S)\alpha\beta}_{\ell HD4} &=&	- \frac{1}{4} \left[ \left( C^{}_{\ell eHD} Y^\dagger_l \right)^{\alpha\beta} + \left( C^{}_{\ell eHD} Y^\dagger_l \right)^{\beta\alpha} \right] \;,
	\nonumber
	\\
	\delta G^{\alpha\beta}_{\ell HD5} &=& -\frac{3}{4} \rmi  \left( C^{}_{\ell eHD} Y^\dagger_l \right)^{\alpha\beta} \;,
	\nonumber
	\\
	\delta G^{(S)\alpha\beta}_{\ell HD6} &=&	- \frac{1}{8} \rmi \left[ \left( C^{}_{\ell eHD} Y^\dagger_l \right)^{\alpha\beta} + \left( C^{}_{\ell eHD} Y^\dagger_l \right)^{\beta\alpha} \right] \;,
	\nonumber
	\\
	\delta G^{\alpha\beta}_{\ell HW} &=& \frac{3}{16} g^{}_2 \left( C^{}_{\ell eHD} Y^\dagger_l \right)^{\alpha\beta} \;,
	\nonumber
	\\
	\delta G^{(A)\alpha\beta}_{\ell HB} &=& \frac{3}{32} g^{}_1 \left[ \left( C^{}_{\ell eHD} Y^\dagger_l \right)^{\alpha\beta} - \left( C^{}_{\ell eHD} Y^\dagger_l \right)^{\beta\alpha} \right] \;,
\end{eqnarray}
where $\delta G^{\cdots}_{\cdots}$ denotes a counterterm in the Green's basis, and an overall factor $1/(16\pi^2 \varepsilon)$ from each counterterm has been omitted. We omit this overall factor from counterterms and wave-function renormalization constants throughout this work. With the help of Eq.~\eqref{eq:counterterm-lehd} and the reduction relations in Ref.~\cite{Zhang:2023kvw}, one can achieve the counterterms for $\Op^{(A)}_{\ell HB}$ and $\Op^{}_{\ell HW}$ in the physical basis,
\begin{eqnarray}\label{eq:counterterm-lehd-phys}
	\delta C^{(A)\alpha\beta}_{\ell HB} &=& \delta G^{(A)\alpha\beta}_{\ell HB} + \frac{1}{16} g^{}_1 \left( \delta G^{\alpha\beta}_{\ell HD2} - \delta G^{\beta\alpha}_{\ell HD2} \right) 
	\nonumber
	\\
	&=& \left[ \frac{3}{32} + \frac{1}{16}  \left( - \frac{3}{2} \right) \right] g^{}_1 \left[ \left( C^{}_{\ell eHD} Y^\dagger_l \right)^{\alpha\beta} - \left( C^{}_{\ell eHD} Y^\dagger_l \right)^{\beta\alpha} \right]
	\nonumber
	\\
	&=& 0   \;,
	\nonumber
	\\
	\delta C^{\alpha\beta}_{\ell HW} &=& \delta G^{\alpha\beta}_{\ell HW} + \frac{1}{4} g^{}_2 \left( \delta G^{(S)\alpha\beta}_{\ell HD4} - \rmi  \delta G^{(S)\alpha\beta}_{\ell HD6} \right) + \frac{1}{16} g^{}_2 \left( \delta G^{\alpha\beta}_{\ell HD2} - \delta G^{\beta\alpha}_{\ell HD2} \right) 
	\nonumber
	\\
	&=& \left[ \frac{3}{16} + \frac{1}{4} \left( -\frac{1}{4} - \frac{1}{8} \right) + \frac{1}{16} \left( - \frac{3}{2} \right) \right] g^{}_2 \left( C^{}_{\ell eHD} Y^\dagger_l \right)^{\alpha\beta}  
	\nonumber
	\\
	&& + \left[ \frac{1}{4} \left( -\frac{1}{4} - \frac{1}{8} \right) + \frac{1}{16} \times \frac{3}{2}  \right] g^{}_2 \left( C^{}_{\ell eHD} Y^\dagger_l \right)^{\beta\alpha} 
	\nonumber
	\\
	&=& 0 \;.
\end{eqnarray}
All results in Eqs.~\eqref{eq:counterterm-lehd} and \eqref{eq:counterterm-lehd-phys} are exactly the same as those calculated by {\sf Matchmakereft}. As apparently revealed by Eq.~\eqref{eq:counterterm-lehd-phys}, contributions to $\delta C^{(A)}_{\ell HB}$ and $\delta C^{}_{\ell HW}$  from $\Op^{}_{\ell e HD}$ cancel each other out and there are no $g\overline{y}$ contributions for $\psi^2H^2X$-$\psi^2H^3D$ entries of the one-loop anomalous dimension matrix, which is in line with the non-renormaliztion theorem.

\subsubsection{Contributions from $\Op^{(S)}_{\ell HD1}$}

As seen from Table~\ref{tab:adm}, the operator $\Op^{(S)}_{\ell HD1}$ makes contribution to the RGEs of all BNC operators. As a result, it is a good choice to recalculate all contributions from $\Op^{(S)}_{\ell HD1}$ as a crosscheck. In this case, one needs to calculate Feynman diagrams with single insertion of $\Op^{(S)}_{\ell HD1}$ and external legs running over field ingredients of all BNC operators to get all UV divergences. Due to a large amount of diagrams, we do not explicitly show them here. Utilizing {\sf FeynRules}, {\sf FeynArts}, and {\sf FeynCalc}, we obtain all counterterms and the wave-function renormalization constants of lepton and Higgs doublets in the Green's basis, which are converted into those in the physical basis, namely,
\begin{eqnarray}\label{eq:wf-lhd1}
	\delta Z^{}_H &=& \frac{1}{4} \left[ \left( 3 - \xi^{}_B \right) g^2_1 + 3 \left( 3 - \xi^{}_W \right) g^2_2 - 4 T \right]  \;,
	\nonumber
	\\
	\delta Z^{}_\ell &=& - \frac{1}{4} \left( \xi^{}_B g^2_1 \mathbbm{1} + 3 \xi^{}_W g^2_2  \mathbbm{1} + 2 Y^{}_l Y^\dagger_l \right) \;,
\end{eqnarray}
and
\begin{eqnarray}\label{eq:counterterm-lhd1}
	\delta C^{(S)\alpha\beta}_{\ell H} &=& - \frac{3}{8} g^2_2  \left(  g^2_2 - 4 \lambda \right) C^{(S)\alpha\beta}_{\ell HD1} + \frac{1}{2} \lambda \left( C^{(S)}_{\ell HD1} Y^{}_l  Y^\dagger_l \right)^{\alpha\beta}  - \frac{1}{2} \left( C^{(S)}_{\ell HD1} Y^{}_l  Y^\dagger_l Y^{}_l Y^\dagger_l  \right)^{\alpha\beta} + \alpha \leftrightarrow \beta \;,
	\nonumber
	\\
	\delta C^{\alpha\beta}_{\ell e HD} &=&  \frac{1}{2} \left( 3g^2_1 - g^2_2 \right) \left( C^{(S)}_{\ell HD1} Y^{}_l \right)^{\alpha\beta} - \left( C^{(S)}_{\ell HD1} Y^{}_l Y^\dagger_l Y^{}_l \right)^{\alpha\beta} \;,
	\nonumber
	\\
	\delta C^{(S)\alpha\beta}_{\ell HD1} &=& -\frac{1}{4} \left[ \xi^{}_B g^2_1 + \left( 3\xi^{}_W - 10 \right) g^2_2 \right] C^{(S)\alpha\beta}_{\ell HD1} + \frac{3}{2} \left( C^{(S)}_{\ell HD1} Y^{}_l Y^\dagger_l \right)^{\alpha\beta} + \alpha \leftrightarrow \beta \;,
	\nonumber
	\\
	\delta C^{(S)\alpha\beta}_{\ell HD2} &=& -2g^2_2 C^{(S)\alpha\beta}_{\ell HD1} - 2 \left( C^{(S)}_{\ell HD1} Y^{}_l Y^\dagger_l \right)^{\alpha\beta} + \alpha \leftrightarrow \beta \;,
	\nonumber
	\\
	\delta G^{\alpha\beta}_{\ell HW} &=& \frac{1}{4} g^3_2 C^{(S)\alpha\beta}_{\ell HD1} - \frac{1}{8} g^{}_2  \left( C^{(S)}_{\ell HD1} Y^{}_l Y^\dagger_l \right)^{\beta\alpha} + \frac{1}{4} g^{}_2  \left( C^{(S)}_{\ell HD1} Y^{}_l Y^\dagger_l \right)^{\alpha\beta} \;,
	\nonumber
	\\
	\delta C^{(A)\alpha\beta}_{\ell HB} &=& - \frac{1}{16} g^{}_1 \left( C^{(S)}_{\ell HD1} Y^{}_l Y^\dagger_l \right)^{\alpha\beta} - \alpha \leftrightarrow \beta \;,
	\nonumber
	\\
	\delta C^{(S)\alpha\beta\gamma\lambda}_{\overline{e}\ell\ell\ell H} &=& \frac{1}{2} \left( g^2_1 - g^2_2 \right) \left[ \left( Y^\dagger_l \right)^{}_{\alpha\beta} C^{(S)\gamma\lambda}_{\ell HD1} +  \left( Y^\dagger_l \right)^{}_{\alpha\gamma} C^{(S)\beta\lambda}_{\ell HD1} +  \left( Y^\dagger_l \right)^{}_{\alpha\lambda} C^{(S)\beta\gamma}_{\ell HD1} \right] 
	\nonumber
	\\
	&& - \frac{1}{4} \left[ \left(C^{(S)}_{\ell HD1} Y^{}_l Y^\dagger_l \right)^{\lambda\beta} \left( Y^\dagger_l \right)^{}_{\alpha\gamma} + \left(C^{(S)}_{\ell HD1} Y^{}_l Y^\dagger_l \right)^{\lambda\gamma} \left( Y^\dagger_l \right)^{}_{\alpha\beta} + \left(C^{(S)}_{\ell HD1} Y^{}_l Y^\dagger_l \right)^{\gamma\beta} \left( Y^\dagger_l \right)^{}_{\alpha\lambda} \right.
	\nonumber
	\\
	&& + \left. \left(C^{(S)}_{\ell HD1} Y^{}_l Y^\dagger_l \right)^{\gamma\lambda} \left( Y^\dagger_l \right)^{}_{\alpha\beta} + \left(C^{(S)}_{\ell HD1} Y^{}_l Y^\dagger_l \right)^{\beta\gamma} \left( Y^\dagger_l \right)^{}_{\alpha\lambda} + \left(C^{(S)}_{\ell HD1} Y^{}_l Y^\dagger_l \right)^{\beta\lambda} \left( Y^\dagger_l \right)^{}_{\alpha\gamma} \right]  \;,
	\nonumber
	\\
	\delta C^{(A)\alpha\beta\gamma\lambda}_{\overline{e}\ell\ell\ell H} &=& \frac{1}{12} \left[  \left( C^{(S)}_{\ell HD1} Y^{}_l Y^\dagger_l \right)^{\lambda\gamma} \left( Y^\dagger_l \right)^{}_{\alpha\beta} -  \left( C^{(S)}_{\ell HD1} Y^{}_l Y^\dagger_l \right)^{\lambda\beta} \left( Y^\dagger_l \right)^{}_{\alpha\gamma} + \left( C^{(S)}_{\ell HD1} Y^{}_l Y^\dagger_l \right)^{\gamma\beta} \left( Y^\dagger_l \right)^{}_{\alpha\lambda} \right.
	\nonumber
	\\
	&& - \left.  \left( C^{(S)}_{\ell HD1} Y^{}_l Y^\dagger_l \right)^{\gamma\lambda} \left( Y^\dagger_l \right)^{}_{\alpha\beta} + \left( C^{(S)}_{\ell HD1} Y^{}_l Y^\dagger_l \right)^{\beta\lambda} \left( Y^\dagger_l \right)^{}_{\alpha\gamma} - \left( C^{(S)}_{\ell HD1} Y^{}_l Y^\dagger_l \right)^{\beta\gamma} \left( Y^\dagger_l \right)^{}_{\alpha\lambda} \right] \;,
	\nonumber
	\\
	\delta C^{(M)\alpha\beta\gamma\lambda}_{\overline{e}\ell\ell\ell H} &=& - \frac{3}{4} \left( g^2_1 + g^2_2 \right) \left[ \left( Y^\dagger_l \right)^{}_{\alpha\beta} C^{(S)\gamma\lambda}_{\ell HD1} - \left( Y^\dagger_l \right)^{}_{\alpha\lambda} C^{(S)\beta\gamma}_{\ell HD1} \right]  - \frac{1}{12} \left[ 4\left( C^{(S)}_{\ell HD1} Y^{}_l Y^\dagger_l \right)^{\lambda\gamma} \left( Y^\dagger_l \right)^{}_{\alpha\beta}  \right.
	\nonumber
	\\
	&& -  \left( C^{(S)}_{\ell HD1} Y^{}_l Y^\dagger_l \right)^{\lambda\beta} \left( Y^\dagger_l \right)^{}_{\alpha\gamma} - 5 \left( C^{(S)}_{\ell HD1} Y^{}_l Y^\dagger_l \right)^{\gamma\beta} \left( Y^\dagger_l \right)^{}_{\alpha\lambda} + 5 \left( C^{(S)}_{\ell HD1} Y^{}_l Y^\dagger_l \right)^{\gamma\lambda} \left( Y^\dagger_l \right)^{}_{\alpha\beta} 
	\nonumber
	\\
	&& + \left. \left( C^{(S)}_{\ell HD1} Y^{}_l Y^\dagger_l \right)^{\beta\lambda} \left( Y^\dagger_l \right)^{}_{\alpha\gamma} - 4 \left( C^{(S)}_{\ell HD1} Y^{}_l Y^\dagger_l \right)^{\beta\gamma} \left( Y^\dagger_l \right)^{}_{\alpha\lambda} \right]  \;,
	\nonumber
	\\
	\delta C^{\alpha\beta\gamma\lambda}_{\overline{d}\ell q \ell H1} &=& - 3 g^2_2 \left( Y^\dagger_{\rm d} \right)^{}_{\alpha\gamma} C^{(S)\beta\lambda}_{\ell HD1} - \frac{3}{2} \left( Y^\dagger_{\rm d} \right)^{}_{\alpha\gamma} \left[ \left( C^{(S)}_{\ell HD1} Y^{}_l Y^\dagger_l \right)^{\beta\lambda} + \left( C^{(S)}_{\ell HD1} Y^{}_l Y^\dagger_l \right)^{\lambda\beta} \right] \;,
	\nonumber
	\\
	\delta C^{\alpha\beta\gamma\lambda}_{\overline{d}\ell q \ell H2} &=& \frac{1}{6} \left( g^2_1 + 9g^2_2 \right) \left( Y^\dagger_{\rm d} \right)^{}_{\alpha\gamma} C^{(S)\beta\lambda}_{\ell HD1} + \frac{1}{2} \left( Y^\dagger_{\rm d} \right)^{}_{\alpha\gamma} \left[ \left( C^{(S)}_{\ell HD1} Y^{}_l Y^\dagger_l \right)^{\beta\lambda} + 2\left( C^{(S)}_{\ell HD1} Y^{}_l Y^\dagger_l \right)^{\lambda\beta} \right] \;,
	\nonumber
	\\
	\delta C^{\alpha\beta\gamma\lambda}_{\overline{d}\ell u e H} &=& \left( Y^\dagger_{\rm d} Y^{}_{\rm u} \right)^{}_{\alpha\gamma} \left( C^{(S)}_{\ell HD1} Y^{}_l \right)^{\beta\lambda}  \;,
	\nonumber
	\\
	\delta C^{\alpha\beta\gamma\lambda}_{\overline{q}u\ell\ell H} &=& \frac{1}{2}  \left(Y^{}_{\rm u} \right)^{}_{\alpha\beta} \left[ 3g^2_2C^{(S)\gamma\lambda}_{\ell HD1} + \left( C^{(S)}_{\ell HD1} Y^{}_l Y^\dagger_l \right)^{\lambda\gamma} + 2 \left( C^{(S)}_{\ell HD1} Y^{}_l Y^\dagger_l \right)^{\gamma\lambda} \right]  \;,
	\nonumber
	\\
	\delta C^{(S)\alpha\beta\gamma\lambda}_{\overline{d}u\ell\ell D} &=&  \left( Y^\dagger_{\rm d} Y^{}_{\rm u} \right)^{}_{\alpha\beta} C^{(S)\gamma\lambda}_{\ell HD1} \;.
\end{eqnarray}
As expected, only $\delta C^{(S)}_{\ell HD1}$ depends on gauge parameters. This gauge-parameter dependence will disappear after the wave-function renormalization constants of lepton and Higgs doublets are taken into account, i.e.,
\begin{eqnarray}\label{eq:counterterm-lhd1-phys}
	\delta C^{(S)\alpha\beta}_{\ell HD1} |^{}_{\rm WF} &=& -\frac{1}{8} \left( 3 g^2_1 - 11 g^2_2 - 4T \right) C^{(S)\alpha\beta}_{\ell HD1} + \frac{7}{4} \left( C^{(S)}_{\ell HD1} Y^{}_l Y^\dagger_l \right)^{\alpha\beta} + \alpha \leftrightarrow \beta \;.
\end{eqnarray}
in which the wave-function renormalization constants of lepton and Higgs doublets in Eq.~\eqref{eq:wf-lhd1} have been exploited. Taking gauge parameters to be $\xi^{}_B = \xi^{}_W = 1$, the results in Eqs.~\eqref{eq:wf-lhd1} and \eqref{eq:counterterm-lhd1} turn out to be fully consistent with those obtained by {\sf Matchmakereft}~\footnote{However, there are some inconsistencies among the results shown in Eqs.~\eqref{eq:counterterm-lhd1} and \eqref{eq:counterterm-lhd1-phys} and those in Ref.~\cite{Liao:2019tep}. We are keeping in contact with the authors to figure out those discrepancies.}.

\section{An Explicit UV Example}\label{sec:UV}

Within the SMEFT, the results in Sec.~\ref{sec:cal} together with those in Ref.~\cite{Zhang:2023kvw} can be exploited to discuss RG running effects on some lepton- and baryon-number-violating processes, such as neutrino masses~\cite{Chala:2021juk}, neutrinoless double beta decay~\cite{Liao:2019tep,Cirigliano:2017djv,Cirigliano:2018yza}, meson decays~\cite{Liao:2019gex,Liao:2020roy}, and nucleon decays~\cite{Beneito:2023xbk}. Thanks to the model-independent property of the SMEFT, those analyses of RG running effects are independent of any UV models and hence only need to be done once. When considering an UV model extending the SM with heavy fields, one only needs to match it onto the SMEFT and substitute the relevant Wilson coefficients in all analyses with their explicit expressions in terms of the UV parameters. For illustration of the relevance of the obtained RGEs for dim-7 operators in Sec.~\ref{sec:cal}, we briefly take the simplified UV model in Ref.~\cite{Beneito:2023xbk} as an example, where the SM is extended with a heavy leptoquark $\omega^{}_2 \left( 3, 1, 2/3 \right) $ and a heavy vector-like fermion $Q^{}_1 \left( 3,2, 1/6 \right) $. The Lagrange of this UV model  is found to be
\begin{eqnarray}\label{eq:UV-Lagrange}
	\mathcal{L}^{}_{\rm UV} &=& \mathcal{L}^{}_{\rm SM} + \left( D^{}_\mu \omega^{}_2 \right)^\dagger \left( D^\mu \omega^{}_2 \right) - M^{}_\omega \omega^\dagger_2 \omega^{}_2 + \overline{Q^{}_1} \rmi \slashed{D} Q^{}_1 - M^{}_Q \overline{Q^{}_1} Q^{}_1 
	\nonumber
	\\
	&& + \left( y^{}_{1,\alpha\beta}  \omega^\dagger_2 \overline{D^{}_{\alpha\rm R}} D^{\rm c}_{\beta\rm R} + y^{}_{2,\alpha} \overline{Q^{}_1} H D^{}_{\alpha \rm R} + y^{}_{3,\alpha} \overline{Q^{}_1} \widetilde{H} U^{}_{\alpha\rm R} + y^{}_{4,\alpha} \omega^{}_2 \overline{Q^{}_1} \ell^{}_{\alpha\rm L} + {\rm h.c.} \right) \;,
\end{eqnarray}
where colour indices are implicitly contracted and the Yukawa coupling $y^{}_1$ is asymmetric, i.e., $y^{}_{1,\beta\alpha} = - y^{}_{1,\alpha\beta}$. The quartic terms of scalar fields are irrelevant and then ignored in Eq.~\eqref{eq:UV-Lagrange}. The first and last terms in the second line of Eq.~\eqref{eq:UV-Lagrange} violate the lepton and baryon numbers by one unit and hence lead to nucleon decays.  Assuming masses of $\omega^{}_2$ and $Q^{}_1$ are roughly in the same order of magnitude, one can integrate out them simultaneously around their mass scale $\Lambda \sim M^{}_\omega \sim M^{}_Q$. At the tree level, the lepton- and baryon-number-violating operators first appear as dimension-seven operators at $\Lambda$, i.e.,
\begin{eqnarray}
	\mathcal{L}^{}_{\rm EFT} \supset C^{(M)\alpha\beta\gamma\lambda}_{\overline{\ell}dddH} \Op^{(M)\alpha\beta\gamma\lambda}_{\overline{\ell}dddH} + C^{\alpha\beta\gamma\lambda}_{\overline{\ell}dud\widetilde{H}} \Op^{\alpha\beta\gamma\lambda}_{\overline{\ell}dud\widetilde{H}} 
\end{eqnarray}
with 
\begin{eqnarray}
	C^{(M)\alpha\beta\gamma\lambda}_{\overline{\ell}dddH} = - \frac{y^\ast_{4,\alpha}}{3M^{}_Q M^2_\omega} \left( 2 y^{}_{2,\beta} y^\dagger_{1,\gamma\lambda} - y^{}_{2,\lambda} y^\dagger_{1,\beta\gamma} - y^{}_{2,\gamma} y^\dagger_{1,\lambda\beta} \right) \;,\quad C^{\alpha\beta\gamma\lambda}_{\overline{\ell}dud\widetilde{H}} = - \frac{2}{M^{}_Q M^2_\omega} y^\ast_{4,\alpha} y^\dagger_{1,\beta\lambda} y^{}_{3,\gamma} \;.\quad
\end{eqnarray}
Due to the asymmetry of $y^{}_1$, only 
\begin{eqnarray}\label{eq:UV-match}
	C^{(M)1112}_{\overline{\ell}dddH} = - C^{(M)1121}_{\overline{\ell}dddH} = -\frac{ y^\dagger_{1,12} y^{}_{2,1} y^\ast_{4,1}}{M^{}_Q M^2_\omega} \quad{\rm and}\quad
	C^{1211}_{\overline{\ell}dud\widetilde{H}} = - C^{1112}_{\overline{\ell}dud\widetilde{H}} = -2 \frac{y^\dagger_{1,21} y^{}_{3,1} y^\ast_{4,1}}{M^{}_Q M^2_\omega}
\end{eqnarray} contribute to nucleon decays. The former one results in the neutron decay $n \to K^+  e^-$ and the latter leads to both neutron and proton decays, i.e., $n \to K^0 \nu$ and $p \to K^+  \nu$. With the help of Eq.~\eqref{eq:psi4H}, one easily gets the RGEs of the above Wilson coefficients, namely
\begin{eqnarray}\label{eq:UV-WCs}
	\dot{C}^{(M)1112}_{\overline{\ell}dddH} &\simeq& - \frac{1}{12} \left( 13 g^2_1 + 27 g^2_2 + 48g^2_3 - 36 y^2_t \right) C^{(M)1112}_{\overline{\ell}dddH} \;,
	\nonumber
	\\
	\dot{C}^{1211}_{\overline{\ell}dud\widetilde{H}} &\simeq& - \frac{1}{12} \left( -23 g^2_1 + 27 g^2_2 + 48g^2_3 - 36 y^2_t \right) C^{1211}_{\overline{\ell}dud\widetilde{H}} \;,
\end{eqnarray}
where the terms suppressed by small Yukawa couplings are omitted. Solving Eq.~\eqref{eq:UV-WCs}, we obtain
\begin{eqnarray}\label{eq:UV-running}
	\frac{C^{(M)1112}_{\overline{\ell}dddH} \left( \Lambda^{}_{\rm EW} \right)}{C^{(M)1112}_{\overline{\ell}dddH} \left( \Lambda\right) } &\simeq& \exp \left( \frac{1}{192\pi^2}  \left( 13 g^2_1 + 27 g^2_2 + 48g^2_3 - 36 y^2_t \right) \ln\frac{\Lambda}{\Lambda^{}_{\rm EW}} \right) \simeq 1.63  \;,
	\nonumber
	\\
	\frac{C^{1211}_{\overline{\ell}dud\widetilde{H}}  \left( \Lambda^{}_{\rm EW} \right)}{C^{1211}_{\overline{\ell}dud\widetilde{H}}  \left( \Lambda\right) } &\simeq& \exp \left( \frac{1}{192\pi^2} \left( -23 g^2_1 + 27 g^2_2 + 48g^2_3 - 36 y^2_t \right)  \ln\frac{\Lambda}{\Lambda^{}_{\rm EW}} \right) \simeq 1.55 
\end{eqnarray}
with $\Lambda^{}_{\rm EW} = M^{}_Z = 91.1876$~GeV and $M^{}_\omega \sim M^{}_Q \sim \Lambda = 10^{10}$~GeV. In Eq.~\eqref{eq:UV-running}, we have ignored the RG running of gauge and top-quark Yukawa couplings for simplicity. It shows that the two Wilson coefficients at the electroweak scale can be roughly enhanced by a factor 1.5 via RG running effects. When low-energy experimental data for nucleon decays, i.e., $p\to K^+ \nu$, $n\to K^0 \nu$ and $n\to K^+ e^-$ is exploited to constrain the UV parameters in Eq.~\eqref{eq:UV-match}, the RG running effects shown in Eq.~\eqref{eq:UV-running} could slightly change the relevant parameter space compared to those without RG running effects. More discussions can be found in Ref.~\cite{Beneito:2023xbk}.

\section{Conclusions}\label{sec:con}

We revisit the renormalization group equations of dim-7 operators in the Standard Model effective field theory in this work. Though such a topic has already been discussed in the previous work~\cite{Liao:2019tep,Liao:2016hru}, the explicit RGEs of dim-7 operators originating from mixing among dim-7 operators themselves are still lacking and we derive them here for the first time. Because we adopt a different strategy to calculate counterterms and perform diagram calculations by means of slightly modified {\sf Matchmakereft} package, our results could provide a crosscheck on previous ones. We also analyse the structure and perturbative power counting of the obtained one-loop anomalous dimension matrix, which are fully consistent with the non-renormalization theorem~\cite{Cheung:2015aba} and the naive dimensional analysis~\cite{Jenkins:2013sda} or power counting rules~\cite{Liao:2017amb}. Furthermore, two partial checks, i.e., contributions to $C^{A}_{\ell HB}$ and $C^{}_{\ell HW}$ from $C^{}_{\ell e HD}$ and all contributions from $C^{(S)}_{\ell HD1}$, are carried out by exploiting {\sf FeynRules}, {\sf FeynArts} and {\sf FeynCalc} with a different quantum field gauge, i.e., the $R^{}_\xi$ gauge, and it turns out that all results exactly match.

Incorporating the results for RGEs of dim-7 operators induced by lower dimensional (i.e., dim-5 and dim-6) operators~\cite{Zhang:2023kvw} into those obtained in this work, we complete the full and explicit RGEs of dim-7 operators in a physical basis shown in Table~\ref{tab:phyb}. One may adopt an other physical basis and get the corresponding RGEs from ours by making some transformations among two bases. Furthermore, with the complete results for dim-7 operators, one can discuss full RG running effects on some appealing lepton- or baryon-number-violating observables or processes as a consequence of dim-7 operators in the SMEFT, such as neutrino masses, neutrinoless double beta decay, meson and nucleon decays.

\section*{Acknowledgements}

We are greatly indebted to Yi Liao and Xiao-dong Ma for useful discussions and for their check and clarification on some discrepancies. We are also grateful to Zhi-zhong Xing for reading this manuscript and giving some important suggestions. This work was supported by the Alexander von Humboldt Foundation.

\begin{appendices}
	
\section{Wave-function Renormalization Constants}\label{app:wf}

The SM wave-function renormalization constants of all fields in the Feynman-t' Hooft gauge $\xi^{}_{B,W,G} = 1$ are given by
\begin{eqnarray}\label{eq:sm-wf}
	\delta Z^{}_B &=& -  \frac{41}{6} g^2_1 \;,
	\nonumber
	\\
	\delta Z^{}_W &=&  \frac{19}{6} g^2_2 \;,
	\nonumber
	\\
	\delta Z^{}_G &=&   7 g^2_3 \;,
	\nonumber
	\\
	\delta Z^{}_H &=&  \frac{1}{2} \left( g^2_1 + 3g^2_2 - 2 T \right)  \;,
	\nonumber
	\\
	\delta Z^{}_\ell &=& - \frac{1}{4} \left( g^2_1 \mathbbm{1} + 3g^2_2  \mathbbm{1} + 2 Y^{}_l Y^\dagger_l \right) \;,
	\nonumber
	\\
	\delta Z^{}_e &=& -  \left( g^2_1 \mathbbm{1} + Y^\dagger_l Y^{}_l \right) \;,
	\nonumber
	\\
	\delta Z^{}_q &=& - \left[ \frac{1}{36} \left( g^2_1 + 27 g^2_2 + 48 g^2_3 \right) \mathbbm{1} + \frac{1}{2} \left( Y^{}_{\rm d} Y^\dagger_{\rm d} + Y^{}_{\rm u} Y^\dagger_{\rm u} \right) \right] \;,
	\nonumber
	\\
	\delta Z^{}_u &=& -  \left[ \frac{4}{9} \left( g^2_1 + 3g^2_3 \right) \mathbbm + Y^\dagger_{\rm u} Y^{}_{\rm u} \right] \;,
	\nonumber
	\\
	\delta Z^{}_d &=& -  \left[ \frac{1}{9} \left( g^2_1 + 12g^2_3 \right) \mathbbm + Y^\dagger_{\rm d} Y^{}_{\rm d} \right] \;.
\end{eqnarray}

\section{Counterterms in the Green's Basis}\label{app:cou}

For convenience, we reproduce the Green's basis for dim-7 operators~\cite{Zhang:2023kvw} in Table~\ref{tab:green-basis}, where $\Opr^{\cdots}_{\cdots}$ denotes to the operator that can be reduced via EoMs. Moreover, we list all results for counterterms of operators in the Green's basis in this appendix. These results can be used to achieve partial contributions to RGEs of higher dimensional operators, and also to perform a check.

\begin{table}[h!]
	\centering
	\renewcommand\arraystretch{1.8}
	\resizebox{\textwidth}{!}{
		\begin{tabular}{l|c|l|c}
			\hline\hline
			\multicolumn{2}{c|}{$\psi^2 H^4$} & \multicolumn{2}{c}{$\psi^4 H$} \\
			\hline
			$\Op^{(S)\alpha\beta}_{\ell H}$ & \makecell[c]{ $\displaystyle\frac{1}{2} \left( \Op^{\alpha\beta}_{\ell H} + \Op^{\beta\alpha}_{\ell H} \right) $ \\ with $\Op^{\alpha\beta}_{\ell H} = \epsilon^{ab} \epsilon^{de}  \left( \ell^a_{\alpha\rm L} C \ell^d_{\beta\rm L} \right) H^b H^e \left( H^\dagger H \right)$ } & $\Op^{(S)\alpha\beta\gamma\lambda}_{\overline{e} \ell\ell\ell H}$ & \makecell[c]{ $\displaystyle\frac{1}{6} \left( \Op^{\alpha\beta\gamma\lambda}_{\overline{e} \ell\ell\ell H} + \Op^{\alpha\lambda\beta\gamma}_{\overline{e} \ell\ell\ell H}  + \Op^{\alpha\gamma\lambda\beta}_{\overline{e} \ell\ell\ell H}  + \Op^{\alpha\beta\lambda\gamma}_{\overline{e} \ell\ell\ell H}  + \Op^{\alpha\gamma\beta\lambda}_{\overline{e} \ell\ell\ell H}  + \Op^{\alpha\lambda\gamma\beta}_{\overline{e} \ell\ell\ell H} \right) $ \\ with $\Op^{\alpha\beta\gamma\lambda}_{\overline{e} \ell\ell\ell H} = \epsilon^{ab} \epsilon^{de} \left( \overline{E^{}_{\alpha\rm R}} \ell^a_{\beta\rm L}\right) \left( \ell^b_{\gamma\rm L} C \ell^d_{\lambda\rm L} \right) H^e$}  \\
			\cline{1-2}
			\multicolumn{2}{c|}{$\psi^2 H^3 D$} & $\Op^{(A)\alpha\beta\gamma\lambda}_{\overline{e} \ell\ell\ell H}$ & $\displaystyle\frac{1}{6} \left( \Op^{\alpha\beta\gamma\lambda}_{\overline{e} \ell\ell\ell H} + \Op^{\alpha\lambda\beta\gamma}_{\overline{e} \ell\ell\ell H}  + \Op^{\alpha\gamma\lambda\beta}_{\overline{e} \ell\ell\ell H}  - \Op^{\alpha\beta\lambda\gamma}_{\overline{e} \ell\ell\ell H}  - \Op^{\alpha\gamma\beta\lambda}_{\overline{e} \ell\ell\ell H}  - \Op^{\alpha\lambda\gamma\beta}_{\overline{e} \ell\ell\ell H} \right) $ \\
			\cline{1-2}
			$\Op^{\alpha\beta}_{\ell e H D}$ & $\epsilon^{ab} \epsilon^{de} \left( \ell^a_{\alpha\rm L} C \gamma^{}_\mu E^{}_{\beta\rm R} \right) H^b H^d {\rm i} D^\mu H^e$ & $\Op^{(M)\alpha\beta\gamma\lambda}_{\overline{e} \ell\ell\ell H}$ &  $\displaystyle\frac{1}{3} \left( \Op^{\alpha\beta\gamma\lambda}_{\overline{e} \ell\ell\ell H} +  \Op^{\alpha\gamma\beta\lambda}_{\overline{e} \ell\ell\ell H} - \Op^{\alpha\lambda\gamma\beta}_{\overline{e} \ell\ell\ell H} - \Op^{\alpha\gamma\lambda\beta}_{\overline{e} \ell\ell\ell H} \right) $ \\
			\cline{1-2}
			\multicolumn{2}{c|}{$\psi^2 H^2 D^2$} & $\Op^{\alpha\beta\gamma\lambda}_{\overline{d} \ell q \ell H 1}$ & $\epsilon^{ab} \epsilon^{de} \left( \overline{D^{}_{\alpha\rm R}} \ell^a_{\beta\rm L} \right) \left( Q^b_{\gamma\rm L} C \ell^d_{\lambda\rm L} \right) H^e $  \\
			\cline{1-2}
			$\Op^{\alpha\beta}_{\ell H D1}$ & $\epsilon^{ab} \epsilon^{de} \left( \ell^a_{\alpha\rm L} C D^\mu \ell^b_{\beta\rm L} \right) H^d D^{}_\mu H^e$ & $\Op^{\alpha\beta\gamma\lambda}_{\overline{d} \ell q \ell H 2}$ & $\epsilon^{ad} \epsilon^{be} \left( \overline{D^{}_{\alpha\rm R}} \ell^a_{\beta\rm L} \right) \left( Q^b_{\gamma\rm L} C \ell^d_{\lambda\rm L} \right) H^e$  \\
			$\Op^{\alpha\beta}_{\ell HD2}$ & $\epsilon^{ad} \epsilon^{be} \left( \ell^a_{\alpha\rm L} C D^\mu \ell^b_{\beta\rm L} \right) H^d D^{}_\mu H^e$ & $\Op^{\alpha\beta\gamma\lambda}_{\overline{d} \ell u e H}$ & $\epsilon^{ab} \left( \overline{D^{}_{\alpha\rm R}} \ell^a_{\beta\rm L} \right) \left( U^{}_{\gamma\rm R} C E^{}_{\lambda\rm R} \right) H^b$ \\
			$\Opr^{(S)\alpha\beta}_{\ell HD3}$ & \makecell[c]{ $\displaystyle\frac{1}{2} \left( \Opr^{\alpha\beta}_{\ell HD3} + \Opr^{\beta\alpha}_{\ell HD3} \right) $ \\ with $\Opr^{\alpha\beta}_{\ell HD3} = \epsilon^{ad} \epsilon^{be} \left( \ell^a_{\alpha\rm L} C  \ell^b_{\beta\rm L} \right) D^\mu H^d D^{}_\mu H^e$} & $\Op^{\alpha\beta\gamma\lambda}_{\overline{q} u \ell \ell H}$ & $\epsilon^{ab} \left( \overline{Q^{}_{\alpha\rm L}} U^{}_{\beta\rm R} \right) \left( \ell^{}_{\gamma\rm L} C \ell^a_{\lambda\rm L} \right) H^b$  \\[0.5cm]
			$\Opr^{(S)\alpha\beta}_{\ell HD4}$ & \makecell[c]{ $\displaystyle\frac{1}{2} \left( \Opr^{\alpha\beta}_{\ell HD4} + \Opr^{\beta\alpha}_{\ell HD4} \right) $ \\ with $\Opr^{\alpha\beta}_{\ell HD4} = \epsilon^{ad} \epsilon^{be} \left( D^\mu \ell^a_{\alpha\rm L} C D^{}_\mu \ell^b_{\beta\rm L} \right) H^d H^e$}  & $\Op^{\alpha\beta\gamma\lambda}_{\overline{\ell} dud \widetilde{H}}$ & $\left( \overline{\ell^{}_{\alpha\rm L}} D^{}_{\beta\rm R} \right) \left( U^{}_{\gamma\rm R} C D^{}_{\lambda\rm R} \right) \widetilde{H} $ \\
			$\Opr^{\alpha\beta}_{\ell HD5}$ & $\epsilon^{ab} \epsilon^{de} \left( \ell^a_{\alpha\rm L} C \sigma^{}_{\mu\nu} D^\mu \ell^b_{\beta\rm L} \right) H^d D^\nu H^e $ & $\Op^{(M)\alpha\beta\gamma\lambda}_{\overline{\ell}dddH}$  & \makecell[c]{ $\displaystyle\frac{1}{3} \left( \Op^{\alpha\beta\gamma\lambda}_{\overline{\ell}dddH} + \Op^{\alpha\gamma\beta\lambda}_{\overline{\ell}dddH} - \Op^{\alpha\beta\lambda\gamma}_{\overline{\ell}dddH} - \Op^{\alpha\lambda\beta\gamma}_{\overline{\ell}dddH} \right) $ \\ with $\Op^{\alpha\beta\gamma\lambda}_{\overline{\ell}dddH} = \left( \overline{\ell^{}_{\alpha\rm L}} D^{}_{\beta\rm R} \right) \left( D^{}_{\gamma\rm R} C D^{}_{\lambda\rm R} \right) H$}  \\[0.5cm]
			$\Opr^{(S)\alpha\beta}_{\ell HD6}$ & \makecell[c]{ $\displaystyle\frac{1}{2} \left( \Opr^{\alpha\beta}_{\ell HD6} + \Opr^{\beta\alpha}_{\ell HD6} \right) $ \\ with $\Opr^{\alpha\beta}_{\ell HD6} = \epsilon^{ad} \epsilon^{be} \left( D^\mu \ell^a_{\alpha\rm L} C \sigma^{}_{\mu\nu} D^\nu \ell^b_{\beta\rm L} \right) H^d H^e$}  & $\Op^{(A)\alpha\beta\gamma\lambda}_{\overline{e}qdd\widetilde{H}}$ & \makecell[c]{ $\displaystyle\frac{1}{2} \left( \Op^{\alpha\beta\gamma\lambda}_{\overline{e}qdd\widetilde{H}} - \Op^{\alpha\beta\lambda\gamma}_{\overline{e}qdd\widetilde{H}} \right) $ \\ with $\Op^{\alpha\beta\gamma\lambda}_{\overline{e}qdd\widetilde{H}} = \epsilon^{ab} \left( \overline{ E^{}_{\alpha\rm R}} Q^a_{\beta\rm L} \right) \left( D^{}_{\gamma\rm R} C D^{}_{\lambda\rm R} \right) \widetilde{H}^b$} \\
			\cline{1-2}
			\multicolumn{2}{c|}{$\psi^2 H^2 X$} & $\Op^{\alpha\beta\gamma\lambda}_{\overline{\ell} d qq \widetilde{H}}$ & $\epsilon^{ab} \left( \overline{\ell^{}_{\alpha\rm L}} D^{}_{\beta\rm R} \right) \left( Q^{}_{\gamma\rm L} C Q^a_{\lambda\rm L} \right) \widetilde{H}^b$   \\
			\cline{1-2}
			$\Op^{(A)\alpha\beta}_{\ell HB}$ & \makecell[c]{ $\displaystyle \frac{1}{2} \left( \Op^{\alpha\beta}_{\ell HB} - \Op^{\beta\alpha}_{\ell HB} \right)$ \\ with $\Op^{}_{\ell HB} = \epsilon^{ab} \epsilon^{de} \left( \ell^a_{\alpha\rm L} C \sigma^{}_{\mu\nu} \ell^d_{\beta\rm L} \right) H^b H^e B^{\mu\nu}$} &   &  \\
			$\Op^{\alpha\beta}_{\ell HW}$ & $\epsilon^{ab} \left( \epsilon \sigma^I \right)^{de} \left( \ell^a_{\alpha\rm L} C \sigma^{}_{\mu\nu} \ell^d_{\beta\rm L} \right) H^b H^e W^{I \mu\nu}$ &  &  \\
			\hline
			\multicolumn{4}{c}{$\psi^4 D$} \\
			\hline
			$\Op^{\alpha\beta\gamma\lambda}_{\overline{e}dddD}$ & $\left(\overline{E^{}_{\alpha\rm R}} \gamma^{}_\mu D^{}_{\beta\rm R} \right) \left( D^{}_{\gamma\rm R} C {\rm i} D^\mu D^{}_{\lambda\rm R} \right) $ & $\Op^{\alpha\beta\gamma\lambda}_{\overline{\ell}qddD}$ & $\left( \overline{\ell^{}_{\alpha\rm L}} \gamma^{}_\mu Q^{}_{\beta\rm L} \right) \left( D^{}_{\gamma\rm R} C {\rm i} D^\mu D^{}_{\lambda\rm  R} \right)$  \\
			$\Op^{\alpha\beta\gamma\lambda}_{\overline{d}u \ell \ell D}$ & $\epsilon^{ab} \left( \overline{D^{}_{\alpha\rm R}} \gamma^{}_\mu U^{}_{\beta\rm R} \right) \left( \ell^a_{\gamma\rm L} C {\rm i} D^\mu \ell^b_{\lambda\rm L} \right)$ & $\Opr^{\alpha\beta\gamma\lambda}_{\overline{\ell}dDqd}$ & $\left( \overline{\ell^{}_{\alpha\rm L}} D^{}_{\beta\rm R} \right) \left( {\rm i}D^\mu Q^{}_{\gamma\rm L} C \gamma^{}_\mu D^{}_{\lambda\rm R} \right)$  \\
			$\Opr^{\alpha\beta\gamma\lambda}_{\overline{d}\ell\ell D u}$ & $\epsilon^{ab} \left( \overline{D^{}_{\alpha\rm R}} \ell^a_{\beta\rm L} \right) \left( \ell^b_{\gamma\rm L }  C \gamma^{}_\mu {\rm i}D^\mu U^{}_{\lambda\rm R } \right) $ & $\Opr^{\alpha\beta\gamma\lambda}_{\overline{\ell} d q D d}$ & $\left( \overline{\ell^{}_{\alpha\rm L}} D^{}_{\beta\rm R} \right) \left( Q^{}_{\gamma\rm L} C \gamma^{}_\mu {\rm i}D^\mu D^{}_{\lambda\rm R} \right)$ \\
			$\Opr^{\alpha\beta\gamma\lambda}_{\overline{d} D \ell\ell u}$ & $\epsilon^{ab} \left( \overline{D^{}_{\alpha\rm R}} {\rm i}D^\mu \ell^a_{\beta\rm L} \right) \left( \ell^b_{\gamma\rm L} C \gamma^{}_\mu U^{}_{\lambda\rm R} \right)$  & & \\
			\hline
			\hline
		\end{tabular}
	}
	\caption{Dimension-7 operators in the Green's basis proposed in Ref.~\cite{Zhang:2023kvw}.}
	\label{tab:green-basis}
\end{table}

\noindent$\bullet~\bm{\psi^2H^4}$
\begin{eqnarray}
	\delta G^{(S)\alpha\beta}_{\ell H} &=& - \frac{1}{4} \left( 3g^2_2 - 40 \lambda \right) C^{(S)\alpha\beta}_{\ell H} - \left( C^{(S)}_{\ell H} Y^{}_l Y^\dagger_l  \right)^{\alpha\beta} + \left( \lambda - \frac{3}{4} g^2_2 \right) \left( C^{}_{\ell e HD} Y^\dagger_l \right)^{\alpha\beta} 
	\nonumber
	\\
	&& + \frac{1}{2} \left( C^{}_{\ell e HD} Y^\dagger_l Y^{}_l Y^\dagger_l \right)^{\alpha\beta} - \frac{3}{8} g^4_2 C^{(S)\alpha\beta}_{\ell HD1} - \lambda \left( C^{(S)}_{\ell HD1} Y^{}_l  Y^\dagger_l \right)^{\alpha\beta} - \frac{1}{2} \left( C^{(S)}_{\ell HD1} Y^{}_l  Y^\dagger_l Y^{}_l Y^\dagger_l  \right)^{\alpha\beta}
	\nonumber
	\\
	&& - \frac{3}{16} \left( g^4_1 + 2 g^2_1 g^2_2 + 3g^4_2 \right) C^{(S)\alpha\beta}_{\ell HD2} - \lambda \left( C^{(S)}_{\ell HD2} Y^{}_l  Y^\dagger_l \right)^{\alpha\beta} - \frac{1}{2} \left( C^{(S)}_{\ell HD2} Y^{}_l  Y^\dagger_l Y^{}_l Y^\dagger_l  \right)^{\alpha\beta}
	\nonumber
	\\
	&& - \frac{3}{2} g^3_2 C^{\alpha\beta}_{\ell HW} - 3 g^{}_2 \left( C^{}_{\ell HW} Y^{}_l Y^\dagger_l \right)^{\alpha\beta} + \frac{3}{2} C^{(S)\gamma\lambda\alpha\beta}_{\overline{e}\ell\ell\ell H} \left( Y^{}_l Y^\dagger_l Y^{}_l \right)^{}_{\lambda\gamma} + C^{(M)\gamma\lambda\alpha\beta}_{\overline{e}\ell\ell\ell H} \left( Y^{}_l Y^\dagger_l Y^{}_l \right)^{}_{\lambda\gamma}
	\nonumber
	\\
	&& + \frac{3}{2} C^{\gamma\alpha\lambda\beta}_{\overline{d}\ell q\ell H1} \left( Y^{}_{\rm d} Y^\dagger_{\rm d} Y^{}_{\rm d} \right)^{}_{\lambda\gamma} - 3 C^{\gamma\lambda\alpha\beta}_{\overline{q}u\ell\ell H} \left( Y^\dagger_{\rm u} Y^{}_{\rm u} Y^\dagger_{\rm u} \right)^{}_{\lambda\gamma} + \alpha \leftrightarrow \beta \;.
\end{eqnarray}

\noindent$\bullet~\bm{\psi^2H^3D}$
\begin{eqnarray}
	\delta G^{\alpha\beta}_{\ell e HD} &=& - \frac{1}{8} \left( 17g^2_1 - 15 g^2_2 -24\lambda \right) C^{\alpha\beta}_{\ell eHD} + \frac{1}{2} \left( Y^{\rm T}_l C^{}_{\ell eHD} Y^\dagger_l \right)^{\beta\alpha} + \frac{3}{2} g^2_1 \left( C^{(S)}_{\ell HD1} Y^{}_l \right)^{\alpha\beta} 
	\nonumber
	\\
	&& + \frac{1}{2} \left( Y^{\rm T}_l C^{(S)}_{\ell HD1} Y^{}_l Y^\dagger_l \right)^{\beta\alpha} + \frac{1}{8} \left( 3g^2_1 - 4\lambda \right) \left( C^{(S)}_{\ell HD2} Y^{}_l \right)^{\alpha\beta} + \frac{1}{4}  \left( Y^{\rm T}_l C^{(S)}_{\ell HD2} Y^{}_l Y^\dagger_l \right)^{\beta\alpha}
	\nonumber
	\\
	&& + \frac{9}{2} g^{}_2 \left( C^{}_{\ell HW} Y^{}_l \right)^{\alpha\beta} - \frac{3}{2} g^{}_2 \left( Y^{\rm T}_l C^{}_{\ell HW} \right)^{\beta\alpha} + 3 g^{}_1 \left( C^{(A)}_{\ell HB} Y^{}_l \right)^{\alpha\beta} - 3 C^{\gamma\alpha\lambda\beta}_{\overline{d}\ell ue H} \left( Y^\dagger_{\rm u} Y^{}_{\rm d} \right)^{}_{\lambda\gamma} \;.
\end{eqnarray}

\noindent$\bullet~\bm{\psi^2H^2D^2}$
\begin{eqnarray}
	\delta G^{\alpha\beta}_{\ell HD1} &=& -\frac{1}{2} g^2_1 C^{(S)\alpha\beta}_{\ell HD1} + \left( C^{(S)}_{\ell HD1} Y^{}_l Y^\dagger_l \right)^{\beta\alpha} - \frac{1}{2} \left( C^{(S)}_{\ell HD1} Y^{}_l Y^\dagger_l \right)^{\alpha\beta} - \frac{1}{16} \left( 21 g^2_1 + 33 g^2_2 + 16\lambda \right) C^{(S)\alpha\beta}_{\ell HD2} 
	\nonumber
	\\
	&& - \frac{1}{2} \left( C^{(S)}_{\ell HD2} Y^{}_l Y^\dagger_l \right)^{\beta\alpha} - \frac{9}{2} g^{}_2 C^{\alpha\beta}_{\ell HW} + \frac{3}{2} g^{}_2 C^{\beta\alpha}_{\ell HW} - 3 g^{}_1 C^{(A)\alpha\beta}_{\ell HB} + \frac{1}{2} \left( 3C^{(S)\gamma\lambda\alpha\beta}_{\overline{e}\ell\ell\ell H} + 3C^{(A)\gamma\lambda\alpha\beta}_{\overline{e}\ell\ell\ell H} \right.
	\nonumber
	\\
	&& + \left. 4 C^{(M)\gamma\lambda\alpha\beta}_{\overline{e}\ell\ell\ell H} + 4 C^{(M)\gamma\alpha\lambda\beta}_{\overline{e}\ell\ell\ell H} \right) \left( Y^{}_l \right)^{}_{\lambda\gamma} + \frac{3}{2} C^{\gamma\alpha\lambda\beta}_{\overline{d}\ell q\ell H1} \left( Y^{}_{\rm d} \right)^{}_{\lambda\gamma} + \frac{3}{2} \left(  C^{\gamma\alpha\lambda\beta}_{\overline{d}\ell q\ell H2} - C^{\gamma\beta\lambda\alpha}_{\overline{d}\ell q\ell H2} \right) \left( Y^{}_{\rm d} \right)^{}_{\lambda\gamma} 
	\nonumber
	\\
	&& - 3 C^{\gamma\lambda\alpha\beta}_{\overline{q}u\ell\ell H} \left( Y^\dagger_{\rm u} \right)^{}_{\lambda\gamma} + 6 C^{(S)\gamma\lambda\alpha\beta}_{\overline{d}u\ell\ell D} \left( Y^\dagger_{\rm u} Y^{}_{\rm d } \right)^{}_{\lambda\gamma} \;,
	\nonumber
	\\
	\delta G^{\alpha\beta}_{\ell HD2} &=& - \frac{3}{2} \left( C^{}_{\ell eHD} Y^\dagger_l \right)^{\alpha\beta} + \frac{3}{2} g^2_2 C^{(S)\alpha\beta}_{\ell HD1} + \frac{3}{2} \left( C^{(S)}_{\ell HD1} Y^{}_l Y^\dagger_l \right)^{\alpha\beta} + \frac{1}{8} \left( 17 g^2_1 + 30 g^2_2 + 16\lambda \right) C^{(S)\alpha\beta}_{\ell HD2} \nonumber
	\\
	&& + \left( C^{(S)}_{\ell HD2} Y^{}_l Y^\dagger_l \right)^{\alpha\beta} - \frac{1}{4} \left( C^{(S)}_{\ell HD2} Y^{}_l Y^\dagger_l \right)^{\beta\alpha} + 3g^{}_2 \left( 3 C^{\alpha\beta}_{\ell HW} - C^{\beta\alpha}_{\ell HW} \right) + 6g^{}_1 C^{(A)\alpha\beta}_{\ell HB} 
	\nonumber
	\\
	&& - \left( 3C^{(S)\gamma\lambda\alpha\beta}_{\overline{e}\ell\ell\ell H} + C^{(M)\gamma\lambda\alpha\beta}_{\overline{e}\ell\ell\ell H} + C^{(M)\gamma\lambda\beta\alpha}_{\overline{e}\ell\ell\ell H} \right) \left( Y^{}_l \right)^{}_{\lambda\gamma} - \frac{3}{2} \left( C^{\gamma\alpha\lambda\beta}_{\overline{d}\ell q \ell H1} + C^{\gamma\beta\lambda\alpha}_{\overline{d}\ell q \ell H1} \right) \left( Y^{}_{\rm d} \right)^{}_{\lambda\gamma} 
	\nonumber
	\\
	&& + 3 \left( C^{\gamma\lambda\alpha\beta}_{\overline{q}u\ell\ell H} + C^{\gamma\lambda\beta\alpha}_{\overline{q}u\ell\ell H} \right) \left( Y^\dagger_{\rm u} \right)^{}_{\lambda\gamma} \;,
	\nonumber
	\\
	\delta G^{(S)\alpha\beta}_{\ell HD3} &=& \frac{3}{4} g^2_2 C^{(S)\alpha\beta}_{\ell HD1} + \frac{3}{4} \left( C^{(S)}_{\ell HD1} Y^{}_l Y^\dagger_l \right)^{\alpha\beta} + \frac{3}{8} g^2_2 C^{(S)\alpha\beta}_{\ell HD2} + \frac{3}{8} \left( C^{(S)}_{\ell HD2} Y^{}_l Y^\dagger_l \right)^{\alpha\beta} 
	\nonumber
	\\
	&& - \frac{1}{4} \left( 3C^{(S)\gamma\lambda\alpha\beta}_{\overline{e}\ell\ell\ell H} + 2C^{(M)\gamma\lambda\alpha\beta}_{\overline{e}\ell\ell\ell H} \right) \left( Y^{}_l \right)^{}_{\lambda\gamma} - \frac{3}{4} C^{\gamma\alpha\lambda\beta}_{\overline{d}\ell q \ell H1} \left( Y^{}_{\rm d} \right)^{}_{\lambda\gamma} + \frac{3}{2} C^{\gamma\lambda\alpha\beta}_{\overline{q}u\ell\ell H} \left( Y^\dagger_{\rm u} \right)^{}_{\lambda\gamma}
	\nonumber
	\\
	&& + \alpha \leftrightarrow \beta \;,
	\nonumber
	\\
	\delta G^{(S)\alpha\beta}_{\ell HD4} &=& - \frac{1}{4} \left( C^{}_{\ell eHD} Y^\dagger_l \right)^{\alpha\beta} + \frac{3}{8} g^2_2 C^{(S)\alpha\beta}_{\ell HD1} + \frac{1}{2} \left( C^{(S)}_{\ell HD1} Y^{}_l Y^\dagger_l \right)^{\alpha\beta} + \frac{3}{16} g^2_2 C^{(S)\alpha\beta}_{\ell HD2} 
	\nonumber
	\\
	&& + \frac{1}{4} \left( C^{(S)}_{\ell HD2} Y^{}_l Y^\dagger_l \right)^{\alpha\beta}  + \frac{3}{2} g^{}_2 C^{\alpha\beta}_{\ell HW} + \alpha \leftrightarrow \beta \;,
	\nonumber
	\\
	\delta G^{\alpha\beta}_{\ell HD5} &=& -\frac{3}{4} \rmi \left( C^{}_{\ell e HD} Y^\dagger_l \right)^{\alpha\beta} - \frac{3}{4} \rmi g^2_2 C^{(S)\alpha\beta}_{\ell HD1} + \frac{1}{4} \rmi \left( C^{(S)}_{\ell HD1} Y^{}_l Y^\dagger_l \right)^{\beta\alpha} + \frac{9}{16} \rmi g^2_2 C^{(S)\alpha\beta}_{\ell HD2} 
	\nonumber
	\\
	&& + \frac{3}{8} \rmi \left( C^{(S)}_{\ell HD2} Y^{}_l Y^\dagger_l \right)^{\alpha\beta} + \frac{1}{2} \rmi \left( C^{(S)}_{\ell HD2} Y^{}_l Y^\dagger_l \right)^{\beta\alpha} \;,
	\nonumber
	\\
	\delta G^{(S)\alpha\beta}_{\ell HD6} &=& - \frac{1}{8} \rmi \left( C^{}_{\ell eHD} Y^\dagger_l \right)^{\alpha\beta} + \frac{1}{4} \rmi g^2_2 C^{(S)\alpha\beta}_{\ell HD1} + \frac{1}{8} \rmi \left( C^{(S)}_{\ell HD1} Y^{}_l Y^\dagger_l \right)^{\alpha\beta} - \frac{1}{32} \rmi \left( g^2_1 - 2g^2_2 \right) C^{(S)\alpha\beta}_{\ell HD2} 
	\nonumber
	\\
	&& + \frac{1}{16} \rmi \left( C^{(S)}_{\ell HD2} Y^{}_l Y^\dagger_l \right)^{\alpha\beta} + \alpha \leftrightarrow \beta \;.
\end{eqnarray}

\noindent$\bullet~\bm{\psi^2H^2X}$
\begin{eqnarray}
	\delta G^{(A)\alpha\beta}_{\ell HB} &=& \frac{3}{32} g^{}_1 \left( C^{}_{\ell eHD} Y^\dagger_l \right)^{\alpha\beta} - \frac{5}{32} g^{}_1 \left( C^{(S)}_{\ell HD1} Y^{}_l Y^\dagger_l \right)^{\alpha\beta}  - \frac{11}{64} g^{}_1 \left( C^{(S)}_{\ell HD2} Y^{}_l Y^\dagger_l \right)^{\alpha\beta} + \frac{3}{4} g^{}_1 g^{}_2 C^{\alpha\beta}_{\ell HW}
	\nonumber
	\\
	&& - \frac{1}{8} \left( 7g^2_2 - 8 \lambda \right) C^{(A)\alpha\beta}_{\ell HB} - \left( C^{(A)}_{\ell HB} Y^{}_l Y^\dagger_l \right)^{\alpha\beta} + \frac{3}{16} g^{}_1 \left( C^{(A)\gamma\lambda\alpha\beta}_{\overline{e}\ell\ell\ell H} - 2 C^{(M)\gamma\lambda\alpha\beta}_{\overline{e}\ell\ell\ell H} \right) \left( Y^{}_l \right)^{}_{\lambda\gamma}
	\nonumber
	\\
	&& - \frac{1}{16} g^{}_1 C^{\gamma\alpha\lambda\beta}_{\overline{d}\ell q\ell H1}\left( Y^{}_{\rm d} \right)^{}_{\lambda\gamma} - \alpha \leftrightarrow \beta \;,
	\nonumber
	\\
	\delta G^{\alpha\beta}_{\ell HW} &=& \frac{3}{16} g^{}_2 \left( C^{}_{\ell e HD} Y^\dagger_\ell \right)^{\alpha\beta} - \frac{1}{16} g^3_2 C^{(S)\alpha\beta}_{\ell HD1} - \frac{3}{16} g^{}_2 \left( C^{(S)}_{\ell HD1} Y^{}_l Y^\dagger_l \right)^{\beta\alpha} + \frac{1}{64} g^{}_2(g^2_1 + 12 g^2_2) C^{(S)\alpha\beta}_{\ell HD2} 
	\nonumber
	\\
	&& + \frac{7}{32} g^{}_2 \left( C^{(S)}_{\ell HD2} Y^{}_l Y^\dagger_l \right)^{\alpha\beta} + \frac{1}{16} g^{}_2 \left( C^{(S)}_{\ell HD2} Y^{}_l Y^\dagger_l \right)^{\beta\alpha} - \frac{1}{8} \left( 6g^2_1 - 15g^2_2 - 16\lambda \right) C^{\alpha\beta}_{\ell HW} 
	\nonumber
	\\
	&& + \frac{17}{8} g^2_2 C^{\beta\alpha}_{\ell HW} + 2\left( C^{}_{\ell HW} Y^{}_l Y^\dagger_l \right)^{\alpha\beta}  + \left( C^{}_{\ell HW} Y^{}_l Y^\dagger_l \right)^{\beta\alpha} - \left( C^{\rm T}_{\ell HW} Y^{}_l Y^\dagger_l \right)^{\beta\alpha} + \frac{3}{4} g^{}_1 g^{}_2 C^{(A)\alpha\beta}_{\ell HB}
	\nonumber
	\\
	&& - \frac{1}{8} g^{}_2 \left( 3 C^{(S)\gamma\lambda\alpha\beta}_{\overline{e}\ell\ell\ell} + C^{(A)\gamma\lambda\alpha\beta}_{\overline{e}\ell\ell\ell} - 2C^{(M)\gamma\lambda\alpha\beta}_{\overline{e}\ell\ell\ell} \right) \left( Y^{}_l \right)^{}_{\lambda\gamma} - \frac{3}{8} g^{}_2 C^{\gamma\beta\lambda\alpha}_{\overline{d}\ell q \ell H1} \left( Y^{}_{\rm d} \right)^{}_{\lambda\gamma} 
	\nonumber
	\\
	&& - \frac{3}{8} g^{}_2 C^{\gamma\alpha\lambda\beta}_{\overline{d}\ell q \ell H2} \left( Y^{}_{\rm d} \right)^{}_{\lambda\gamma} - \frac{3}{8} g^{}_2 C^{\gamma\beta\lambda\alpha}_{\overline{d}\ell q \ell H2} \left( Y^{}_{\rm d} \right)^{}_{\lambda\gamma} \;.
\end{eqnarray}

\noindent$\bullet~\bm{\psi^4H}$


\noindent$\bullet~\bm{\psi^4D}$
\begin{eqnarray}
	\delta G^{\alpha\beta\gamma\lambda}_{\overline{e}dddD} &=& 2\left( 2C^{(M)\rho\beta\gamma\lambda}_{\overline{\ell}dddH} - C^{(M)\rho\gamma\beta\lambda}_{\overline{\ell}dddH} \right) \left( Y^\dagger_l \right)^{}_{\alpha\rho} + 2 C^{(A)\alpha\rho\beta\gamma}_{\overline{e}qdd\widetilde{H}} \left( Y^{}_{\rm d} \right)^{}_{\rho\lambda} -g^2_1 C^{(S)\alpha\beta\gamma\lambda}_{\overline{e}dddD} 
	\nonumber
	\\
	&& - C^{(S)\rho\sigma\gamma\lambda}_{\overline{\ell}qddD} \left( Y^\dagger_l \right)^{}_{\alpha\rho} \left( Y^{}_{\rm d} \right)^{}_{\sigma\beta} \;,
	\nonumber
	\\
	\delta G^{\alpha\beta\gamma\lambda}_{\overline{d}u\ell\ell D} &=& \frac{1}{2} \left( 2 C^{(S)\gamma\lambda}_{\ell HD1} + C^{(S)\gamma\lambda}_{\ell HD2} \right) \left( Y^\dagger_{\rm d} Y^{}_{\rm u} \right)^{}_{\alpha\beta} + \frac{1}{2} C^{\alpha\gamma\beta\rho}_{\overline{d}\ell ue H} \left( Y^\dagger_l \right)^{}_{\rho\lambda}  - \frac{1}{2} \left( C^{\rho\beta\gamma\lambda}_{\overline{q}u\ell\ell H} - C^{\rho\beta\lambda\gamma}_{\overline{q}u\ell\ell H} \right) \left( Y^\dagger_{\rm d} \right)^{}_{\alpha\rho}
	\nonumber
	\\
	&& - \frac{1}{18} \left( 11g^2_1 + 27g^2_2 + 24g^2_3 \right) C^{(S)\alpha\beta\gamma\lambda}_{\overline{d}u\ell\ell D} \;,
	\nonumber
	\\
	\delta G^{\alpha\beta\gamma\lambda}_{\overline{d}D\ell\ell u} &=& \frac{1}{6} \left( g^2_1 + 9g^2_2 \right) C^{(S)\alpha\lambda\beta\gamma}_{\overline{d}u\ell\ell D} - \frac{1}{2} C^{\alpha\gamma\lambda\rho}_{\overline{d}\ell u eH} \left( Y^\dagger_l \right)^{}_{\rho\beta} \;,
	\nonumber
	\\
	\delta G^{\alpha\beta\gamma\lambda}_{\overline{d}\ell\ell Du} &=& \frac{1}{3} g^2_1 C^{(S)\alpha\lambda\beta\gamma}_{\overline{d}u\ell\ell D} + \frac{1}{2} \left( C^{\rho\lambda\beta\gamma}_{\overline{q}u\ell\ell H} - C^{\rho\lambda\gamma\beta}_{\overline{q}u\ell\ell H} \right) \left( Y^\dagger_{\rm d} \right)^{}_{\alpha\rho} + \frac{1}{2} \left( C^{\alpha\beta\rho\gamma}_{\overline{d}\ell q\ell H1} + 2 C^{\alpha\beta\rho\gamma}_{\overline{d}\ell q\ell H2} \right) \left( Y^{}_{\rm u} \right)^{}_{\rho\lambda} \;,
	\nonumber
	\\
	\delta G^{\alpha\beta\gamma\lambda}_{\overline{\ell}qddD} &=& - C^{(A)\rho\beta\gamma\lambda}_{\overline{e}qdd\widetilde{H}} \left( Y^{}_l \right)^{}_{\alpha\rho} - \frac{3}{2} C^{(S)\rho\sigma\gamma\lambda}_{\overline{e}dddD} \left( Y^{}_l \right)^{}_{\alpha\rho} \left( Y^\dagger_{\rm d} \right)^{}_{\sigma\beta} - \frac{1}{36} \left( g^2_1 + 27 g^2_2 + 48g^2_3 \right) C^{(S)\alpha\beta\gamma\lambda}_{\overline{\ell}qddD} 
	\nonumber
	\\
	&& - \frac{1}{2} C^{(S)\alpha\rho\sigma\lambda}_{\overline{\ell}qddD} \left( Y^\dagger_{\rm d} \right)^{}_{\sigma\beta} \left( Y^{}_{\rm d} \right)^{}_{\rho\gamma} - \frac{1}{2} C^{(S)\alpha\rho\gamma\sigma}_{\overline{\ell}qddD} \left( Y^\dagger_{\rm d} \right)^{}_{\sigma\beta} \left( Y^{}_{\rm d} \right)^{}_{\rho\lambda} \;,
	\nonumber
	\\
	\delta G^{\alpha\beta\gamma\lambda}_{\overline{\ell}dqDd} &=& - \frac{1}{2} \left(  2 C^{\alpha\beta\gamma\rho}_{\overline{\ell}dqq\widetilde{H}} - C^{\alpha\beta\rho\gamma}_{\overline{\ell}dqq\widetilde{H}} \right) \left( Y^{}_{\rm d} \right)^{}_{\rho\lambda} - \frac{3}{2} C^{(S)\rho\beta\sigma\lambda}_{\overline{e}dddD} \left( Y^{}_l \right)^{}_{\alpha\rho} \left( Y^\dagger_{\rm d} \right)^{}_{\sigma\gamma} + \frac{4}{9} \left( g^2_1 + 3g^2_3\right) C^{(S)\alpha\gamma\beta\lambda}_{\overline{\ell}qddD} \;,
	\nonumber
	\\
	\delta G^{\alpha\beta\gamma\lambda}_{\overline{\ell}dDqd} &=& - \frac{1}{2} C^{\alpha\beta\rho\lambda}_{\overline{\ell}dud\widetilde{H}} \left( Y^\dagger_{\rm u} \right)^{}_{\rho\gamma} + \left( 2 C^{(M)\alpha\rho\lambda\beta}_{\overline{\ell}dddH} - C^{(M)\alpha\lambda\rho\beta}_{\overline{\ell}dddH} \right) \left( Y^\dagger_{\rm d} \right)^{}_{\rho\gamma} + C^{(A)\rho\gamma\beta\lambda}_{\overline{e}qdd\widetilde{H}} \left( Y^{}_l \right)^{}_{\alpha\rho} 
	\nonumber
	\\
	&& + \frac{1}{2} C^{(S)\rho\beta\lambda\sigma}_{\overline{e}dddD} \left( Y^{}_l \right)^{}_{\alpha\rho} \left( Y^\dagger_{\rm d} \right)^{}_{\sigma\gamma} + \frac{1}{3} g^2_1 C^{(S)\alpha\gamma\beta\lambda}_{\overline{\ell}qddD} - C^{(S)\alpha\rho\lambda\sigma}_{\overline{\ell}qddD} \left( Y^{}_{\rm d} \right)^{}_{\rho\beta} \left( Y^\dagger_{\rm d} \right)^{}_{\sigma\gamma} 
	\nonumber
	\\
	&& + \frac{1}{2} C^{(S)\alpha\rho\beta\sigma}_{\overline{\ell}qddD} \left( Y^{}_{\rm d} \right)^{}_{\rho\lambda} \left( Y^\dagger_{\rm d} \right)^{}_{\sigma\gamma} \;.
\end{eqnarray}

\end{appendices}

\end{document}